# Magnon spectroscopy in the electron microscope

Demie Kepaptsoglou[1,2,3*†], José Ángel Castellanos-Reyes[4†], Adam Kerrigan[2,3], Júlio Alves do Nascimento[2,3], Paul M. Zeiger[4], Khalil El hajraoui[1,2], Juan Carlos Idrobo[5,6], Budhika G. Mendis[7], Anders Bergman[4], Vlado K. Lazarov[2,3], Ján Rusz[4*], and Quentin M. Ramasse[1,8,9*]

[1]*SuperSTEM Laboratory, SciTech Daresbury Campus, Daresbury, WA4 4AD, UK.*

[2]*School of Physics, Engineering and Technology, University of York, Heslington, YO10 5DD, UK.*

[3]*JEOL NanoCentre, University of York, Heslington, YO10 5DD, UK.*

[4]*Department of Physics and Astronomy, Uppsala University, Box 516, Uppsala, 75120, Sweden.*

[5]*Materials Science and Engineering Department, University of Washington, Seattle, WA 98195, USA.*

[6]*Physical and Computational Sciences Directorate, Pacific Northwest National Laboratory, Richland, WA 99354, USA.*

[7]*Department of Physics, Durham University, Durham, DH1 3LE, UK.*

[8]*School of Chemical and Process Engineering, University of Leeds, Leeds, LS2 9JT, UK.*

[9]*School of Physics and Astronomy, University of Leeds, Leeds, LS2 9JT, UK.*

*Corresponding author(s). E-mail(s): dmkepap@superstem.org; jan.rusz@physics.uu.se; qmramasse@superstem.org.

[†]These authors contributed equally to this work.

## Summary

The miniaturisation of transistors is approaching its limits due to challenges in heat management and information transfer speed [1]. To overcome these obstacles, emerging technologies such as spintronics [2] are being developed, which leverage the electron's spin in addition to its charge. Local phenomena at interfaces or structural defects will greatly influence the efficiency of spin-based devices, making the ability to study spin-wave propagation at the nano- and atomic scales a key challenge [3, 4]. The development of high-spatial-resolution tools to probe spin waves, also called magnons, at relevant length-scales is thus essential to understand how their properties are affected by local features. Here, we detect bulk THz magnons at the nanoscale using scanning transmission electron microscopy. By employing high-resolution electron energy-loss spectroscopy with hybrid-pixel electron detectors, we overcome the challenges posed by weak signals to map THz-magnon excitations in a thin NiO nanocrystal. Advanced inelastic electron scattering simulations corroborate our findings. These results open new avenues for detecting magnons, exploring their dispersions and their modifications arising from nanoscale structural or chemical defects. This marks a milestone in magnonics and presents exciting opportunities for the development of spintronic devices.

## Main Text

The use of spin currents in information transistors is predicted to offer non-volatility, faster data processing, higher integration densities, and lower power consumption [5–7], thanks to degrees of freedom emerging in quantum materials that exhibit unique spin-dependent properties, including topologically protected spin states [5]. At the same time, recent developments in antiferromagnetic (AFM) spintronics have demonstrated that AFM materials, such as NiO, provide a promising platform for spin injection and transport in the THz domain [8], and spin-torque control [9]. Magnons, the collective excitations of the spin lattice in ferro- and antiferro-magnets, which can be visualised semi-classically as a wave of precessing magnetic moments [10], are becoming a cornerstone of quantum technology [11] through proposed spin-current-based device architectures. Of crucial importance are localised phenomena at the nano- and atomic scales such as the scattering of spin waves at hetero-interfaces and structural defects in materials, which can affect spin injection, spin-wave transmission, spin-torque switching, and spin-to-charge conversion. As a result, the ability to achieve (sub-)nm-resolution magnon detection is considered one of the main challenges in the field of magnonics [3, 4].

Electrons as a probe for magnon excitations are commonly used through surface-scattering of low-energy electrons in reflection high-resolution electron energy-loss spectroscopy (HREELS), using either spin- or non-polarised electron sources [10, 12–14]. While HREELS can probe the energy-momentum dispersion of magnons with high energy resolution, it is limited to the study of surface excitations of ultra-thin films over large length-scales due to limitations in the scattering cross-sections, spatial resolution, and penetration depth intrinsic to the technique. Similarly, other experimental approaches widely used to study magnons at high energy and momentum resolutions, such as inelastic neutron scattering [15], time-resolved Kerr microscopy [16], or Brillouin light-scattering [17], are also fundamentally limited to spatial resolutions of hundreds of nanometres and often to large sample volumes. Consequently, magnon information from nanometre-sized features, such as defects and buried interfaces, is not accessible. New approaches in vector magnetometry using nitrogen-vacancy ($NV^-$) centres sensing have recently shown great promise in mapping surface magnetic textures and detecting magnons at the nanometre-scale with high sensitivity [18, 19], although the fabrication of $NV^-$-point-defect sensors remain challenging.

Since its first demonstration [20, 21], meV-level (vibrational) electron-energy-loss spectroscopy (EELS) in a scanning transmission electron microscope (STEM) has been developing at a swift pace. Several key experimental milestones have been achieved: the detection of atomic-level contrast in vibrational signals [22], the spectral signature of individual impurity atoms [23], spatially resolved measurements on point- and line-defects in crystalline materials [24, 25] as well as momentum-resolved measurements using nanoscale beams [26, 27]. With energy losses due to magnon excitations occupying the same spectral window as phonons, ranging from a few to a few hundred meV in solid-state materials [10, 12–14], the promise of detecting magnons in an electron microscope is exciting from both fundamental research and applications points of view.

Recent theoretical studies on inelastic magnon scattering in an electron microscope confirmed the detectability of magnons as diffuse inelastic scattering, and demonstrated the tantalising prospect of obtaining atomically localised magnon information [28–30]. This exploratory work also highlighted experimental challenges such as the separation of phonon from magnon diffuse scattering, as the latter is predicted to be several orders of magnitude weaker than the former and yet occupy a similar energy-loss span. Here we tackle the challenges of detecting magnons at the nanoscale using high-resolution EELS in the STEM, and we present the first direct detection of magnons with STEM-EELS. Furthermore, we demonstrate that at the interface between a NiO thin film and a non-magnetic substrate, the magnon signal is exclusively confined within the film, confirming that magnons can be mapped with nanometre spatial resolution. We show that, although challenging, the detection of the inherently weak magnon signal is possible, thanks in part to the dynamic range of hybrid-pixel electron detectors [31, 32]. The experiments are supported by state-of-the-art numerical simulations of electron scattering [33], underpinned by atomistic spin-dynamics (ASD) simulations [34].

The primary challenge in achieving magnon-EELS is that the energy ranges of phonon and magnon losses overlap, with the weaker magnon signal likely to be overshadowed by the inherently stronger lattice vibration modes. However, these types of losses follow different dispersion relations [35] and, thus, should be differentiable in momentum-resolved experiments, given a suitable choice of material whose magnon and phonon branches are sufficiently separated in momentum and energy. For this purpose we have selected NiO as a model system; in addition to being of interest for spin-transfer-based devices [36], its dispersion relations of phonon and magnon modes have been shown to meet our requirements of momentum and energy separation in the THz range [37, 38].

A schematic representation of the experimental geometry used for the momentum-resolved experiments is presented in Fig. 1a. The instrument's electron optics are adjusted to a low convergence angle to form a diffraction-limited nanometre-size electron probe, while achieving sufficient momentum resolution (Methods). The electron probe is kept stationary on a region of interest of the NiO crystal, across which the zone-axis orientation is perfectly maintained, estimated to extend no further than a few nanometres from the nominal probe position. The stability of the microscope sample stage, with typical drift measured below 0.5 nm per hour, enables hours-long acquisitions on a nanometre-sized area of the NiO crystal (Methods), a timescale still far shorter than necessary for magnon spectroscopy experiments using, e.g., inelastic neutron scattering, where days of integration from bulk samples can be required. A narrow rectangular (slot) collection aperture for EELS is employed for the angle-resolved measurements [39]. The slot aperture is aligned to select a row of systematic Bragg reflections (Fig. 1b) and to produce two-dimensional intensity maps of energy loss (or frequency) $\omega$ versus momentum transfer $\mathbf{q}$.

Figure 2 shows examples of such measurements acquired along the 220 and 002 rows of systematic Bragg reflections in NiO (corresponding to the $\Gamma \rightarrow$ M and $\Gamma \rightarrow$ X $\mathbf{q}$-paths, respectively). Figures 2a,b are the as-acquired intensity maps of energy versus momentum transfer ($\omega$-$\mathbf{q}$ maps). The $\omega$-$\mathbf{q}$ maps show two intense bands dispersing around 30 and 50 meV, which correspond to the NiO longitudinal-acoustic (LA) and longitudinal-optical (LO) phonon branches, respectively, in agreement with previous experimental [37] and theoretical work [38, 40]. These are labelled on Extended Data Fig. 2, where gain LA phonon branches are also visible (higher energy-gain branches are outside of the recorded energy window).

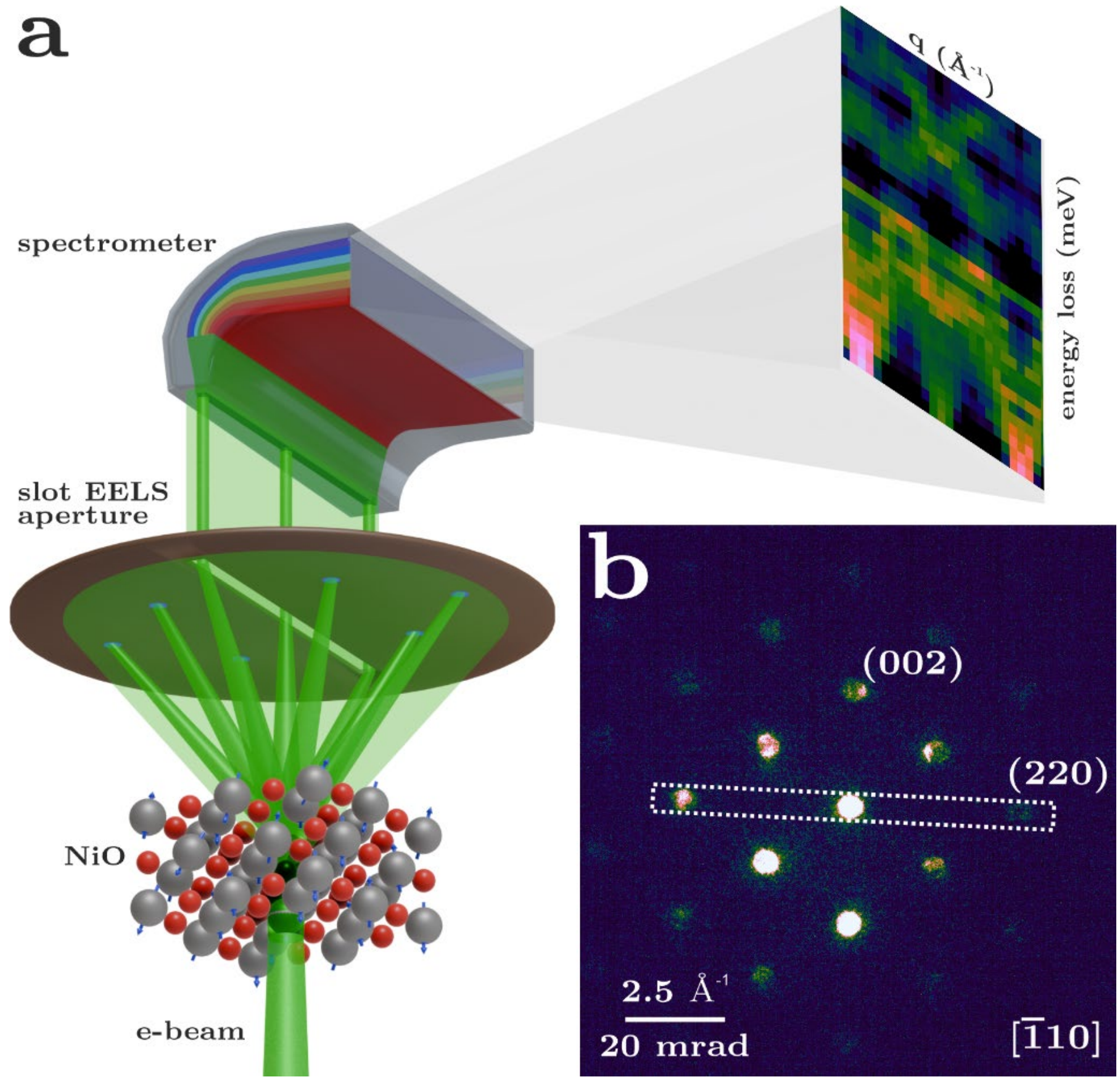

**Fig. 1 | Experimental geometry of momentum-resolved EELS.** ***a.*** *Schematic representation of the geometry of* $\omega$-**q** *vibrational EELS measurements using a rectangular slot collection aperture.* ***b.*** *Experimental diffraction pattern along the NiO* $[\bar{1}10]$ *zone axis at a 2.25 mrad convergence angle, with the monochromating slit inserted, showing the orientations of the slot EELS collection aperture along the 220 row of systematic Bragg reflections in diffraction space.*

Since the intensity of magnon-EELS is expected to be significantly lower than that of phonons [28, 30], and given that magnon modes in NiO occur at higher energy losses compared to phonons, scaling the data by the square of the energy loss (intensity × $E^2$) provides a useful means to enhance the visibility of weaker features above the decaying zero-loss-peak tail in the meV range, while avoiding possible errors in background fitting: see Supplementary note 1. The scaled but otherwise unprocessed $\omega$-**q** maps, presented in Supplementary Fig. 1b,e, readily show additional spectral bands above 80 meV along Γ → M and Γ → X, present but hard to discern in the non-scaled data.

Subtracting the background from the tail of the phonon signal using a first-order log-polynomial model offers a clear illustration of the dispersion of the two spectral bands above 80 meV, in Fig. 2c and 2d for the Γ → M and Γ → X directions (with signal-to-noise ratios, SNRs, of 22 and 59, respectively; see Supplementary Note 3 for details on SNR estimation). Supplementary Note 2 explores the robustness of different background models.

In Fig. 2c, two lobes of spectral intensity are visible on either side of the M point, with a maximum peak at 100 meV. The lobes' intensity is asymmetric around the M point, with a stronger signal between Γ and M compared to the lobe between M and Γ′. The intensity tends to 0 towards each of the Brillouin zone vertices. The same asymmetry is also evident in the 002 data, Fig. 2d, but the separation between lobes is less pronounced along the shorter Γ→X distances due to limited momentum resolution and spectral smearing caused by averaging. Supplementary Fig. 3 shows how better branch separation can be achieved with less averaging at the cost of higher noise and lower SNR, as discussed in Supplementary Note 3.

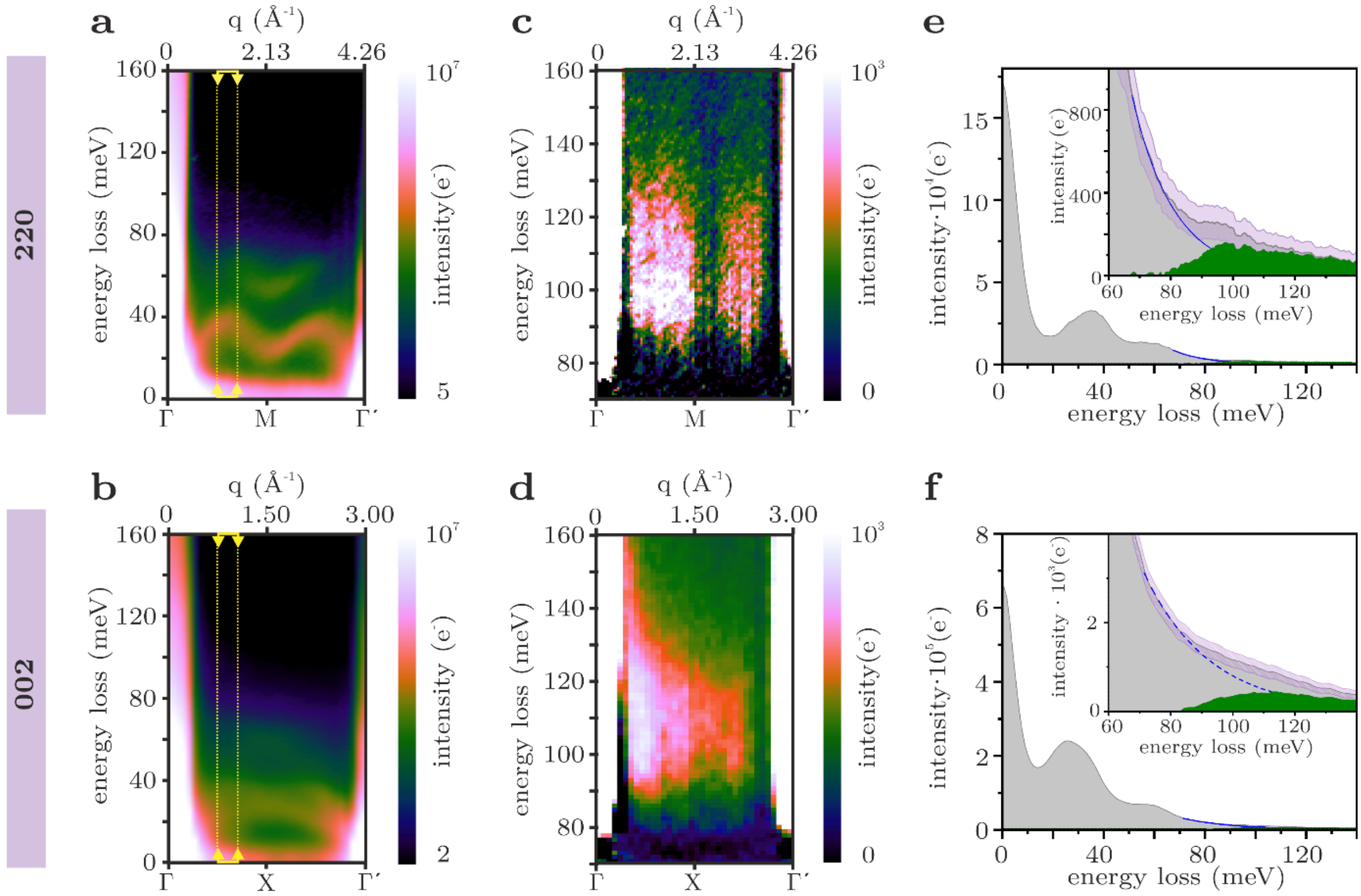

**Fig. 2 | Momentum-resolved vibrational EELS measurements of NiO.** ***a,b.*** *As-acquired ω-***q** *maps along the 220 and 002 rows of reflections, respectively, showing the dispersion of the NiO LA and LO phonon branches. For clarity the intensity of the maps (calibrated in electrons, e⁻) is displayed on a logarithmic colour scale.* ***c,d.*** *Background-subtracted ω-***q** *maps showing the dispersion of the magnon bands at ~100 meV.* ***e,f.*** *Integrated spectra at the momentum positions marked by arrows in panels a,b. Inset: background-subtracted spectra (green shaded area), obtained by removing a first-order log-polynomial model (dotted blue line) from the raw signal (shaded grey area). Pink shaded areas illustrate the error bars as confidence bands at a +/-5σ level (σ is the standard deviation, calculated by assuming the magnon scattering and noise populations are Poisson distributed; see Supplementary Note 3).*

The presence of these bands in the non-scaled data is confirmed in Fig. 2e by the integrated intensity profiles of the $\omega$-**q** map over a narrow momentum window ($\Delta q$ = 0.22 Å$^{-1}$, to avoid spectral broadening through momentum averaging) at q = 1.24 Å$^{-1}$, the value at which the intensity is maximum along Γ → M, marked by yellow arrows and dashed lines in Fig. 2a. This highlights the shape of the peak, with a rising edge from 80 meV reaching a maximum at ~100 meV, before a weaker feature extending up to 120 meV. Similarly in the integrated intensity profile of the Γ → X **q**-path in Fig. 2f, the band at 100 meV is observed at q = 0.97 Å$^{-1}$ (over a $\Delta q$ = 0.2 Å$^{-1}$ window, Fig. 2d), comparatively further away from the Γ point than the band observed in the Γ → M direction. This signal is unambiguously above the experimental error, as illustrated in Fig. 2e,f by confidence bands (pink shaded areas) at a 5σ-confidence-level (σ is the standard deviation, calculated by assuming the magnon scattering and noise populations are Poisson distributed: see Supplementary Note 3).

The observed spectral bands emerge in the same energy-momentum space where magnon modes are expected for NiO. The magnon density of states is known to shift to higher energies with decreasing temperature [41], so the acquisition temperature difference explains a ~20 meV blue-shift between EELS (room temperature) and neutron experiments (10 K in ref. [37]), a conclusion borne out by simulations discussed below. Furthermore, the asymmetry of detected bands along both **q**-paths, including their appearing most prominently above background further away from Γ in the Γ → X direction than along Γ → M, is consistent with neutron-scattering experiments [37, 41]. Here, the asymmetry is also likely due in part to the lower intensity away from the direct beam.

Due to the overlap of the elastic (or ZLP), phonon and magnon signals along **q** it is difficult to quantify the absolute intensity of the observed bands we attribute to magnons. Nevertheless, an estimate can be given by the integral of the signal within the Γ → M → Γ' window (excluding Γ points); after background subtraction (Fig. 2c), the integrated magnon intensity is estimated as ~$8.5\times10^{4}$ e$^{-}$. For comparison, the corresponding integrated phonon intensity (for all branches) over the same $\Delta q$ is three orders of magnitude higher, ~$2.0\times10^{7}$ e$^{-}$, while the total integrated intensity across the slot aperture, assumed to be representative of the total beam intensity impinging on the sample, was $5.2\times10^{9}$ e$^{-}$, in agreement with expected relative intensities of the magnon and total scattered EELS signals [28]. A similar intensity analysis holds for the Γ → X data.

To support our experimental findings, we have performed first-principles calculations of momentum-resolved phonon and magnon EELS using parameters that reflect the experimental conditions, in particular the sample temperature, magnetic environment, and the microscope's electron-optical parameters (Methods). Figure 3 summarises the results of the simulations for both experimental **q**-paths. We observe phonon EELS bands reaching up to about 70 meV for both **q**-paths, with a small gap around 40 meV (Fig. 3a,c). This matches the two dominant features observed in the experimental momentum-resolved phonon EELS datasets, albeit with a slightly higher energy-loss gap at 50 meV. Due to the long acquisition times required to reveal the emergence of the magnon bands, and the chosen balance between momentum and spatial resolutions, some spectral smearing obscures the finer details of the experimental phonon dispersion. For completeness, a better-resolved $\omega$-**q** map is presented in Extended Data Fig. 2 comparing very favourably with the calculated phonon dispersion in Fig. 3a (but in which the magnon signal is fainter due to shorter integration), and data accumulation is discussed in Supplementary note 3. The rich pattern observed in phonon simulations contrasts with magnon simulations, which display negligible fractional scattering intensities at energy losses below 60 meV, with two broad-yet-well-isolated peaks at energy losses between 80-120 meV (Figs. 3b,d).

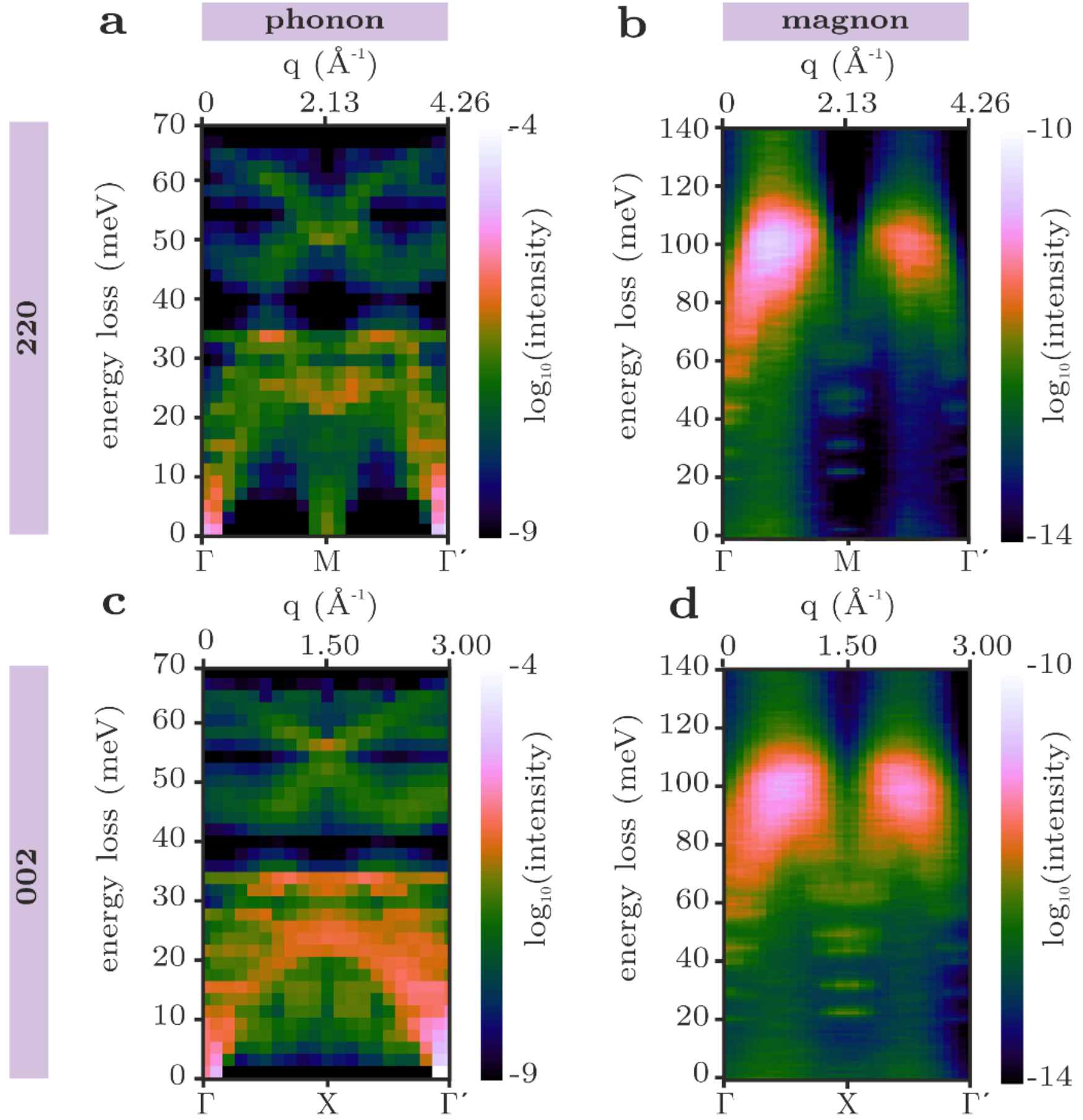

**Fig. 3 | Calculated vibrational and magnon EELS of NiO.** *Simulated momentum-resolved EELS (dispersion curves) of* ***(a,c)*** *phonon and* ***(b,d)*** *magnon excitations, along the* $\Gamma \rightarrow M$ *and* $\Gamma \rightarrow X$ ***q****-paths of the Brillouin zone for NiO, respectively. Experimental parameters such as sample temperature, environmental magnetic field inside the microscope and electron-optical parameters are considered (Methods).*

The theoretical calculations closely match the features emerging above the phonon EELS background observed in experiments, Fig. 2. The maximum simulated magnon EELS intensity appears around 100 meV at room temperature, as observed experimentally, whereas simulations of the magnon dispersion at 10 K (Extended Data Fig. 3) predict an energy-loss peak close to 120 meV, as expected from neutron experiments at this temperature. The measured asymmetry along the momentum axis, also present in neutron-scattering data, is well reproduced, showing a considerably higher intensity for the magnon EELS peak closer to the Γ point in the centre of the diffraction plane, when compared to the peak near the 220 or 002 Γ' points, respectively. This is intrinsic to the magnon EELS scattering strength, although, as discussed above, it is exacerbated experimentally due to the lower intensity away from the direct STEM-EELS beam. Similarly, along the momentum direction, the calculated magnon intensity maximum appears comparatively closer to the X point than to the M point, as in the experiments. Furthermore, the intensity peak shape in the energy-loss direction matches the experimental observations, with an extended spectral tail towards 120 meV in the Γ → M calculations, compared to a slightly more rounded peak centred at 100 meV in the Γ → X case. Theoretical and experimental integrated spectra and dispersions are displayed side-by-side in Extended Data Fig. 4, and a broadened version of the simulations to mimic the finite experimental resolution is shown in Supplementary Fig. 4. These illustrate the excellent match, and confirm the interpretation of the experimental intensities around 100 meV as being the spectral signature of scattering arising from magnon excitations in the nanometre-sized NiO crystal.

A key advantage of STEM-EELS is its unique ability to probe spectral signals at high spatial resolution. Spatially resolved measurements across a 30 nm thick film of NiO grown atop an yttrium-stabilised zirconia (YSZ) substrate demonstrate the potential of this technique for mapping magnons at the nanometre scale. Figure 4 illustrates a dark-field EELS experiment [22], whereby an angstrom-sized electron probe is rastered across the sample, the EELS signal being collected by a circular aperture displaced off the optical axis (Methods). This geometry was suggested in earlier work to be favourable for spatially resolved magnon detection [28]. An averaged spectrum shows a well-defined magnon peak at 100 meV (Fig. 4e), while, remarkably, the integrated map (Fig. 4d) shows that the signal is confined within the NiO film, disappearing completely within a sub-nm distance of the interface with the (non-magnetic) YSZ substrate and into the hole above the film surface. Spatially resolved spectral variations are explored in Supplementary Note 6.

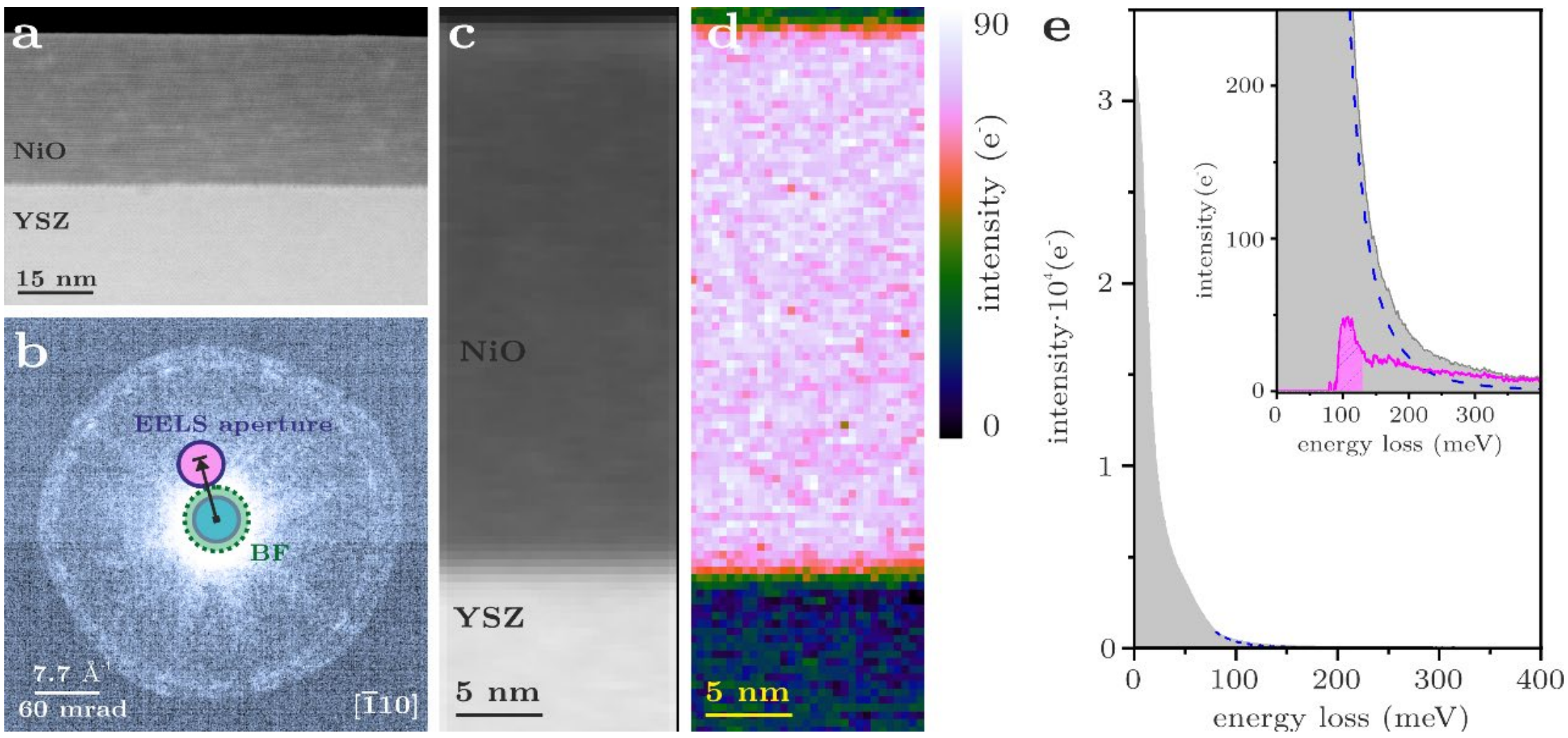

**Fig. 4 | Spatially resolved magnon EELS measurements across a NiO thin film.** ***a.*** *High-angle annular-dark-field Image of a 30 nm NiO thin film grown on an yttrium-stabilised zirconia (YSZ) substrate.* ***b.*** *Experimental diffraction pattern along the NiO* $[\bar{1}10]$ *zone axis at a 31 mrad convergence angle, with the monochromating slit inserted, showing the off-axis displacement of the round EELS collection aperture along the [002] direction (marked by the black arrow).* ***c.*** *Asymmetric (displaced) annular-dark-field Image (a-ADF) acquired during EELS measurements.* ***d.*** *Integrated intensity map of the magnon peak over the energy window indicated by a shaded area, after background subtraction, as illustrated in* ***(e)*** *with the signal integrated over the whole NiO film area (the background-subtracted version is presented, pink curve, inset).*

Inelastic magnon simulations for a thin slab of NiO are consistent with the experiments, and predict a smaller intensity of the magnon signal at the slab's edges compared to its centre, with a broader, more featureless post-peak tail (Supplementary Fig. 7 and 8). The thorough interpretation of signal differences near heterointerfaces would require the development of appropriate modelling frameworks, and the observed drop in signal intensity observed near the interface with YSZ may have other causes (e.g., beam propagation). Nevertheless, this experiment demonstrates unambiguously the nanometre-scale mapping of magnons in an electron microscope.

## Discussion and Conclusions

Our observation of magnon excitations in STEM-EELS arrives a decade after the milestone report of the detection of phonons in an electron microscope [20]. One may reasonably expect further developments of magnon EELS to follow a similar trajectory and pace as its phonon counterpart, with a blueprint for studies of magnon dispersions and their nano-scale modifications in the vicinity of surfaces, interfaces or defects. These will create a radical new way of studying magnetism and spin-wave excitations at the length-scales relevant for device fabrication, using widely-applied numerical analysis techniques, e.g. to extract exchange interaction parameters, using spatially resolved experimental data. In this context, we emphasise that in addition to the spatially resolved magnon map in a thin film (Fig. 4), the $\omega$-$\mathbf{q}$ measurements were performed with a probe diameter smaller than 2 nm, with a 10 nm wide region of interest at most, across which the NiO single-crystal was perfectly on axis to enable a careful probing of the chosen diffraction directions, as illustrated by images of the area used for the experiments (Extended Data Fig. 1). This combination of high spatial resolution with the flexibility to optimize the acquisition geometry for dispersion measurements when sample size, heterogeneity, or spatial selectivity are important, is unique to STEM [26]. Although momentum resolution is limited in the nanoprobe regime, balancing momentum and spatial resolutions will be necessary to understand the underlying scattering physics and spin-wave propagation in nano-scale objects.

The significantly lower cross-section for scattering of fast electrons from magnons compared to phonons [28, 29, 33] remains a challenge. In our experiments, calculations suggest that we can expect to detect one electron scattered by magnons only every 1-2 seconds, while there can be thousands of phonon-scattered electrons every second under equivalent settings. This may limit the ability to differentiate magnon dispersions from phonon branches present within the same region of $\omega$-$\mathbf{q}$ space, placing constraints on the choice of materials systems for which the direct observation of magnon peaks will be possible in the immediate future.

However, several systems of interest for magnonics, including metals [28, 35] and oxides [42], present favourable separation between phonons and magnon branches, with relatively high magnon scattering cross-sections. There are therefore many materials to explore as the technique develops, where low detection efficiency and spectral overlaps are not prohibitive.

Furthermore, alternative strategies can be deployed to enable a more effective, albeit less direct detection of magnons in STEM-EELS in cases for which magnons and phonon branches are close. The interaction of magnons with other excitations in the same energy range can give rise to other quasiparticles such as magnon polarons (the hybridised state between phonons and magnons), which offer less-challenging spectroscopic fingerprints. These interactions are characterised by modifications of the dispersion behaviour (e.g. the appearance of spectral band anti-crossings) [43, 44]. Preliminary reports suggest this strategy may be promising [45, 46]. Similarly, the different dependence of magnons on external sample stimuli (e.g., temperature) compared to phonons may provide an efficient means to disentangle the magnon signal from higher intensity features [30].

Finally, on-going technological and methodological developments in STEM-EELS are opening experimental avenues. Electron counting detectors enable measurements hampered only by Poisson noise [31]. Although comparatively shorter experiments are an advantage for the STEM-EELS approach, the exceptional stability of STEM instruments with low sample drift, operating at reduced acceleration voltages minimizing sample damage combined with advanced data acquisition procedures such as multi-frame recording [47], make it possible in principle to significantly increase acquisition times, approaching day-long timescales that are common in techniques such as angle-resolved photo-emission (ARPES) [48] or inelastic neutron scattering, while retaining nano- or atomic-scale spatial resolutions. Longer acquisition times will provide an efficient mitigation strategy to circumvent the low magnon-EELS signal, leading to higher-fidelity detection at high spatial resolution. With the availability of tools like variable-field pole pieces [49] and liquid-helium-temperature stages [50], along with newer-generations monochromators, future magnon STEM-EELS experiments will gain new controls over magnetic field and extended temperature ranges. These can be used to suppress magnon or phonon modes or to enhance energy separation, marking a new era for magnonics studies at the nanometre scale.

## References

[1] Mamaluy, D. & Gao, X. The fundamental downscaling limit of field effect transistors. *Applied Physics Letters* **106**, 193503 (2015). https://doi.org/10.1063/1.4919871.

[2] Wolf, S. A. *et al.* Spintronics: A spin-based electronics vision for the future. *Science* **294**, 1488–1495 (2001). https://www.science.org/doi/abs/10.1126/science.1065389.

[3] Barman, A. *et al.* The 2021 Magnonics Roadmap. *Journal of Physics: Condensed Matter* **33**, 413001 (2021). https://iopscience.iop.org/article/10.1088/1361-648X/abec1a.

[4] Flebus, B. *et al.* The 2024 magnonics roadmap. *Journal of Physics: Condensed Matter* **36**, 363501 (2024). https://dx.doi.org/10.1088/1361-648X/ad399c.

[5] Chumak, A. V., Serga, A. A. & Hillebrands, B. Magnonic crystals for data processing. *Journal of Physics D: Applied Physics* **50**, 244001 (2017). https://dx.doi.org/10.1088/1361-6463/aa6a65.

[6] Puebla, J., Kim, J., Kondou, K. & Otani, Y. Spintronic devices for energy-efficient data storage and energy harvesting. *Communications Materials* **1**, 24 (2020). https://doi.org/10.1038/s43246-020-0022-5.

[7] Han, W., Otani, Y. & Maekawa, S. Quantum materials for spin and charge conversion. *npj Quantum Materials* **3**, 27 (2018). https://www.nature.com/articles/s41535-018-0100-9.

[8] Rezende, S. M., Azevedo, A. & Rodríguez-Suárez, R. L. Introduction to anti-ferromagnetic magnons. *Journal of Applied Physics* **126**, 151101 (2019). https://doi.org/10.1063/1.5109132.

[9] Moriyama, T., Oda, K., Ohkochi, T., Kimata, M. & Ono, T. Spin torque control of antiferromagnetic moments in NiO. *Scientific Reports* **8**, 14167 (2018). https://doi.org/10.1038/s41598-018-32508-w.

[10] Vollmer, R., Etzkorn, M., Anil Kumar, P., Ibach, H. & Kirschner, J. Spin-polarized electron energy loss spectroscopy: a method to measure magnon energies. *Journal of Magnetism and Magnetic Materials* **272-276**, 2126–2130 (2004). https://doi.org/10.1016/j.jmmm.2003.12.506.

[11] Flebus, B., Rezende, S. M., Grundler, D. & Barman, A. Recent advances in magnonics. *Journal of Applied Physics* **133**, 160401 (2023). https://doi.org/10.1063/5.0153424.

[12] Zakeri, K., Zhang, Y. & Kirschner, J. Surface magnons probed by spin-polarized electron energy loss spectroscopy. *Journal of Electron Spectroscopy and Related Phenomena* **189**, 157–163 (2013). https://doi.org/10.1016/j.elspec.2012.06.009.

[13] Rajeswari, J. *et al.* Surface spin waves of fcc cobalt films on Cu(100): High-resolution spectra and comparison to theory. *Physical Review B* **86**, 165436 (2012). https://doi.org/10.1103/PhysRevB.86.165436.

[14] Ibach, H., Bocquet, F. C., Sforzini, J., Soubatch, S. & Tautz, F. S. Electron energy loss spectroscopy

with parallel readout of energy and momentum. *Review of Scientific Instruments* **88**, 033903 (2017). https://doi.org/10.1063/1.4977529.

[15] Shirane, G., Minkiewicz, V. J. & Nathans, R. Spin Waves in 3d Metals. *Journal of Applied Physics* **39**, 383–390 (1968). https://doi.org/10.1063/1.2163453.

[16] Keatley, P. S. *et al.* A platform for time-resolved scanning Kerr microscopy in the near-field. *Review of Scientific Instruments* **88**, 123708 (2017). https://doi.org/10.1063/1.4998016.

[17] Sebastian, T., Schultheiss, K., Obry, B., Hillebrands, B. & Schultheiss, H. Micro- focused Brillouin light scattering: imaging spin waves at the nanoscale. *Frontiers in Physics* **3**, 35 (2015). https://www.frontiersin.org/journals/physics/articles/10.3389/fphy.2015.00035.

[18] Hu, Z., He, Z. Wang, Q., Chou, C. -T., Hou, J. T., and Liu, L., Nonlinear Magnetic Sensing with Hybrid Nitrogen-Vacancy/Magnon Systems, *Nano Letters* **24**, 15731–15737 (2024). https://doi.org/10.1021/acs.nanolett.4c04459.

[19] Xu, Y., Zhang, W., and Tian, C. Recent advances on applications of $NV^-$ magnetometry in condensed matter physics. *Phononics Research* **11**, 393-412 (2023). https://doi.org/10.1364/PRJ.471266.

[20] Krivanek, O. L. *et al.* Vibrational spectroscopy in the electron microscope. *Nature* **514**, 209 (2014). https://doi.org/10.1038/nature13870.

[21] Miyata, T. *et al.* Measurement of vibrational spectrum of liquid using monochromated scanning transmission electron microscopy-electron energy loss spectroscopy. *Microscopy* **63**, 377 (2014). https://doi.org/10.1093/jmicro/dfu023.

[22] Hage, F. S., Kepaptsoglou, D. M., Ramasse, Q. M. & Allen, L. J. Phonon Spectroscopy at Atomic Resolution. *Physical Review Letters* **122**, 016103 (2019). https://link.aps.org/doi/10.1103/PhysRevLett.122.016103.

[23] Hage, F. S., Radtke, G., Kepaptsoglou, D. M., Lazzeri, M. & Ramasse, Q. M. Single-atom vibrational spectroscopy in the scanning transmission electron micro- scope. *Science* **367**, 1124-1127 (2020). https://doi.org/10.1126/science.aba1136.

[24] Yan, X. *et al.* Single-defect phonons imaged by electron microscopy. *Nature* **589**, 65–69 (2021). https://doi.org/10.1038/s41586-020-03049-y.

[25] Hoglund, E. R. *et al.* Emergent interface vibrational structure of oxide superlattices. *Nature* **601**, 556–561 (2022). https://www.nature.com/articles/s41586-021-04238-z.

[26] Hage, F. S. *et al.* Nanoscale momentum-resolved vibrational spectroscopy. *Science Advances* **4**, eaar7495 (2018). https://www.science.org/doi/10.1126/sciadv.aar7495.

[27] Senga, R. *et al.* Position and momentum mapping of vibrations in graphene nanostructures. *Nature* **573**, 247–250 (2019). https://www.nature.com/articles/s41586-019-1477-8.

[28] Lyon, K. *et al.* Theory of magnon diffuse scattering in scanning transmission electron microscopy. *Physical Review B* **104**, 214418 (2021). https://doi.org/10.1103/PhysRevB.104.214418.

[29] Mendis, B. Quantum theory of magnon excitation by high energy electron beams. *Ultramicroscopy* **239**, 113548 (2022). https://doi.org/10.1016/j.ultramic.2022.113548.

[30] Castellanos-Reyes, J. Á. *et al.* Unveiling the impact of temperature on magnon diffuse scattering detection in the transmission electron microscope. *Physical Review B* **108**, 134435 (2023). https://link.aps.org/doi/10.1103/PhysRevB.108.134435.

[31] Plotkin-Swing, B. *et al.* Hybrid pixel direct detector for electron energy loss spectroscopy. *Ultramicroscopy* **217**, 113067 (2020). https://doi.org/10.1016/j.ultramic.2020.113067.

[32] Fernandez-Perez, S. *et al.* Characterization of a hybrid pixel counting detector using a silicon sensor and the IBEX readout ASIC for electron detection. *Journal of Instrumentation* **16**, P10034 (2021). https://dx.doi.org/10.1088/1748-0221/16/10/P10034.

[33] Castellanos-Reyes, J. Á., Zeiger, P. & Rusz, J. Dynamical theory of angle-resolved electron energy loss and gain spectroscopies of phonons and magnons in transmission electron microscopy including multiple scattering effects. *Physical Review Letters* **134***,* 036402 (2025). https://doi.org/10.1103/PhysRevLett.134.036402.

[34] The Uppsala atomistic spin dynamics code, UppASD. https://github.com/UppASD/UppASD (2023). Last accessed 2023-11-13.

[35] Wu, X., Liu, Z. & Luo, T. Magnon and phonon dispersion, lifetime, and thermal conductivity of iron from spin-lattice dynamics simulations. *Journal of Applied Physics* **123**, 85109 (2018). https://doi.org/10.1063/1.5020611.

[36] Baldrati, L. *et al.* Spin transport in multilayer systems with fully epitaxial NiO thin films. *Physical Review B* **98**, 14409 (2018). https://doi.org/10.1103/PhysRevB.98.014409.

[37] Sun, Q. *et al.* Mutual spin-phonon driving effects and phonon eigenvector renormalization in nickel (II) oxide. *Proceedings of the National Academy of Sciences* **119**, e2120553119 (2022). https://pnas.org/doi/full/10.1073/pnas.2120553119.

[38] Betto, D. *et al.* Three-dimensional dispersion of spin waves measured in NiO by resonant inelastic x-ray scattering. *Physical Review B* **96**, 020409 (2017). https://doi.org/10.1103/PhysRevB.96.020409.

[39] Qi, R. *et al.* Four-dimensional vibrational spectroscopy for nanoscale mapping of phonon dispersion in BN nanotubes. *Nature Communications* **12**, 1179 (2021). https://doi.org/10.1038/s41467-021-21452-5.

[40] Gupta, B. R. K. & Verma, M. P. Application of three body force shell model to the lattice dynamics

of transition metal oxides MnO, CoO and NiO. *Journal of Physics and Chemistry of Solids* **38**, 929–932 (1977). https://doi.org/10.1016/0022-3697(77)90133-0.

[41] Woo, C. H., Wen, H., Semenov, A. A., Dudarev, S. L. & Ma, P.-W. Quantum heat bath for spin-lattice dynamics. *Physical Review B* **91**, 104306 (2015). https://doi.org/10.1103/PhysRevB.91.104306.

[42] Princep, A. J., Ewings, R. A., Ward, S., Toth, S., Dubs, C., Prabhakaran, D., and Boothroyd, A. T. The full magnon spectrum of yttrium iron garnet. *npj Quantum Materials* **2**, 63 (2017). https://doi.org/10.5286/isis.e.73945114.

[43] Godejohann, F. *et al.* Magnon polaron formed by selectively coupled coherent magnon and phonon modes of a surface patterned ferromagnet. *Physical Review B* **102**, 144438 (2020). https://link.aps.org/doi/10.1103/PhysRevB.102.144438.

[44] Olsson, K. S. *et al.* Spin-phonon interaction in yttrium iron garnet. *Physical Review B* **104**, L020401 (2021). https://link.aps.org/doi/10.1103/PhysRevB.104.L020401.

[45] Reifsnyder, A. *et al.* Detecting Magnon-Phonon Coupling in the Scanning Transmission Electron Microscope. *Microscopy and Microanalysis* **30**, ozae044.772 (2024). https://doi.org/10.1093/mam/ozae044.772.

[46] El hajraoui, K. *et al.* Towards the In-situ Detection of Spin Charge Accumulation at a Metal/Insulator Interface Using STEM-EELS Technique. *Microscopy and Microanalysis* **28**, 2338–2339 (2022). https://doi.org/10.1017/S1431927622008972.

[47] Jones, L. *et al.* Optimising multi-frame ADF-STEM for high-precision atomic- resolution strain mapping. *Ultramicroscopy* **179**, 57–62 (2017). https://doi.org/10.1016/j.ultramic.2017.04.007.

[48] Iwasawa, H. *et al.* Accurate and efficient data acquisition methods for high-resolution angle-resolved photoemission microscopy. *Scientific Reports* **8**, 17431 (2018). https://doi.org/10.1038/s41598-018-34894-7.

[49] Shibata, N. *et al.* Atomic-resolution electron microscopy in a magnetic field free environment. *Nature Communications* **10**, 2308 (2019). https://doi.org/10.1038/s41467-019-10281-2.

[50] Rennich, E., Sung, S. H., Agarwal, N., Hovden, R. & El Baggari, I. Liquid Helium TEM Sample Holder with High Stability and Long Hold Times. *Microscopy and Microanalysis* **29**, 1696–1697 (2023). https://doi.org/10.1093/micmic/ozad067.874.

## Methods

### *Materials and Specimen Preparation*

Electron microscopy specimens were prepared conventionally from a commercially available NiO single crystal substrate (Surface Preparation Laboratory) by crushing (pestle and mortar), dispersing in chloroform and drop casting on standard lacey carbon support grids. Additional specimens were prepared by focused ion beam (FIB), using a Hitachi NX5000 Ethos triple-beam instrument, from a 30 nm-thick NiO thin film grown on an yttrium-stabilised zirconia (YSZ) substrate by molecular beam epitaxy. Sample thickness was estimated using the standard log-ratio method from low-loss EELS spectra to be approximately 30-40 nm in the areas used for all data presented here. Grains with a suitable zone-axis orientation were specifically chosen so their diffraction patterns aligned close to the main axis of the slot aperture for momentum-resolved acquisitions. Small rotational adjustments were carried out using the microscope's projector lens system, to ensure perfect alignment of the row of diffraction spots to the slot axis. When this was not possible, the sample was removed from the microscope and rotated to match the measured rotation of pre-screened grains of interest. The effective regions of interest across which the crystal areas were both thin enough, overhanging any lacey carbon support material, and perfectly on zone axis (judged from the symmetry of the observed intensities in the diffraction pattern), were estimated to be no larger than 10 nm across in most cases.

### *High Resolution Electron Energy Loss Spectroscopy and data processing*

High-resolution electron energy loss spectroscopy (EELS) measurements were performed in a Nion UltraSTEM 100MC "Hermes" aberration-corrected dedicated scanning transmission electron microscope. The instrument is equipped with a Nion-designed high-resolution ground-potential monochromator, a Nion Iris high-energy-resolution energy-loss spectrometer and a Dectris ELA hybrid-pixel direct electron detector optimized for EELS at low acceleration voltages [51]. The instrument was operated at 60 kV acceleration voltage to minimise sample damage and take advantage of higher inelastic cross-section for EELS experiments [52], with the electron optics adjusted to achieve a 2.25 mrad convergence angle corresponding to a diffraction limited ∼1.3 nm electron probe diameter (Extended Data Fig. 1). Although smaller convergence angles can be achieved, this choice offered an adequate balance between probe size, momentum resolution and electron optics stability. The nominal energy resolution was 7.2 meV, set by the position of the monochromator energy-selection slit and estimated by the full-width at half-maximum (FWHM) of the zero-loss peak (ZLP) in vacuum (Extended Data Fig. 2). Some energy broadening occurs with the beam traversing the sample and with data accumulation (see Supplementary Note 3). The resulting probe current after monochromation and with the optical parameters chosen to ensure the highest achievable energy resolution was ∼1 pA. For the momentum-resolved experiments, a rectangular (slot) aperture (0.125 mm × 2 mm) was employed, with the projector optics adjusted to match the full angular size of the beam to the width of the slit. The momentum resolution is dictated by the combined effect of the convergence and collection apertures used. This is often described using an 'effective' collection angle, the quadratic sum of the convergence half-angle $\alpha$ and collection half angle β [53]. To be exact, the momentum resolution can be expressed as $k\sqrt{\sin^2\alpha + \sin^2\beta}$ where $k$ is the electron beam wavenumber, and $\alpha$ and $\beta$ are the convergence and collection semi-angles, respectively. In the case of a rectangular momentum-selecting EELS slit, which is here chosen to have a width in the energy-dispersive direction that matches the beam convergence, it is easy to see that the momentum resolution can for simplicity be considered to be limited by the size of the diffraction spots on the EELS camera (that is, the full beam convergence angle). Defining the electron wavenumber as $2\pi/\lambda$ for a wavelength $\lambda = 0.0487$ Å at 60 kV acceleration voltage, this yields a momentum resolution of $\Delta q = 0.4$ Å$^{-1}$, consistent with previous nanoscale momentum-resolved experiments using a similar set of parameters [26]. The slot aspect-ratio allows an angular range in the momentum direction of 50 mrad. The pixel size along the momentum direction was 0.057 Å$^{-1}$/pixel and 0.088 Å$^{-1}$/pixel for the $\Gamma \rightarrow$ M and $\Gamma \rightarrow$ X directions, respectively. The spectrometer dispersion was set to 1.0 or 0.5 meV/channel (002 and 220 acquisitions, respectively). Each $\omega$-**q** dataset was a multi-frame acquisition of the whole 2-dimensional extent of the spectrometer camera. A typical dataset comprised 15,000 frames, using an exposure time of 75 ms per frame, resulting in an acquisition time of 22 min per single dataset (limited only by the data-export capabilities of the microscope operating software). The multi-frame stacks were subsequently assessed for energy drift, aligned using rigid image registration, and integrated. The final data presented here are sums of multiple such integrated datasets acquired consecutively (necessary as explained above due to data-size-handling limitations) from the same specimen area, on the same day and under identical experimental conditions. No other post-processing was applied. The impact of this averaging on practical energy and momentum resolution is explored in Supplementary Note 3.

Specifically, the dataset in Fig. 2a is the sum of 90,000 individual camera frames, acquired along the $[\bar{1}10]$ zone axis of NiO (Fig. 1b) with a dwell time of 75 ms, corresponding to a total of 2 h acquisition time. Due to the integration and averaging of multiple datasets, some spectral smearing results in an effective energy resolution for the experiment of 9.2 meV, measured by the full-width at half-maximum (FWHM) of the zero-loss peak (ZLP) in the final integrated dataset. The dataset in Fig. 2b is the sum of 60,000 individual camera frames, acquired along the [100] zone axis of NiO (Extended Data Figure 1) with a dwell time of 75 ms, corresponding to a total of 1.2 h acquisition time. The effective energy resolution of the experiment was 11 meV, measured by the full-width at half-maximum (FWHM) of the zero-loss peak (ZLP) in the integrated dataset. The smaller number of frames required to obtain comparable SNR in the case of the 002 data (Fig. 2b) is due to a tip change occurring between the two sets of experiments, resulting in higher probe currents available at otherwise similar electron optical settings.

The reproducibility of the results was thoroughly assessed through the acquisition of numerous independent datasets, across several days of experiments and on various areas of the samples, all of which exhibited the same spectral features. For completeness, an additional dataset is presented in Supplementary Fig. 5. The likelihood of spurious carbon-related vibrational modes affecting the results (due to adventitious carbon build-up or neighbouring support film) was fully excluded, as spectra recorded on an area of protective carbon on the FIB-prepared sample showed an entirely different energy loss and spectral fingerprint, as illustrated in Supplementary Fig. 6 and associated discussion.

Spatially resolved high-resolution EELS measurements were performed in the dark-field (off-axis) geometry [20], using a 31 mrad convergence semi-angle corresponding to ~1 Å electron probe diameter (Fig. 4). The EELS collection semi-angle was 22 mrad with the spectrometer entrance aperture displaced by 55 mrad (or 7 $Å^{-1}$ momentum transfer) along the [002] direction in momentum space. The energy resolution was 9 meV, as estimated by the full-width at half-maximum of the zero-loss peak (ZLP) in vacuum, and the spectrometer dispersion was set to 1.0 meV/channel with a camera dwell time of 15 ms per pixel. The data presented in Fig. 4**c-e** is the sum of eight multi-pass spectrum images (each containing 10 spectrum image frames, for a total of 80 frames) with the electron beam rastered over a 15 × 40 nm area. Each spectrum image was aligned for ZLP and spatial drift before summing all the frames [48]. For the map presented in Fig. 4**d** the spectral data was denoised using Principal Component Analysis, as implemented in Gatan's GMS 3.6 software suite. The intensity map of the magnon signal was generated from the integrated spectral intensity range of 80-130 meV after background subtraction using a first-order log-polynomial background to model the decaying intensity tail underlying the preceding phonon signal. A detailed analysis of observed spatial variations of the spectral signal is provided in Supplementary Note 6 and Supplementary Fig. 6, including a comparison of spectral signatures of the NiO thin film, the YSZ substrate and the carbon-based protective sample cap, confirming the distinct and unique fingerprint of the magnon signal.

***Theoretical calculations***

The study of phonon and magnon excitations through EELS requires computational methods that accurately capture complex physical effects, such as multiple scattering and dynamical diffraction, while addressing the specific challenges posed by different materials systems. To meet these requirements, we employ two distinct approaches: the TACAW and the FRFPMS methods. These are described below with an overview of their physics, implementation, and application in this work.

The **Time-Autocorrelation of Auxiliary Wavefunctions (TACAW)** method [33] is a novel and versatile framework for simulating angle-resolved EELS of both phonons and magnons. It works by Fourier transforming the temperature-dependent time auto-correlation of the auxiliary electron beam wave-function. For phonons, TACAW relies on frozen phonon multislice simulations [54] using molecular dynamics to generate atomic displacement snapshots. For magnons, it employs frozen magnon multislice simulations [28], incorporating atomistic spin dynamics to model magnetic excitations. This method captures multiple scattering, thermal, and dynamical diffraction effects, enabling the resolution of both energy-loss and energy-gain processes.

The **Frequency-Resolved Frozen Phonon Multislice (FRFPMS)** method [55], on the other hand, focuses specifically on phonon excitations. In this work, it utilizes snapshots of atomic displacements derived from density functional theory (DFT) simulations to represent vibrational modes. These snapshots are grouped into frequency bins for detailed spectral decomposition. While TACAW could not be applied to phonons in this case due to the absence of an accurate molecular dynamics (MD) potential for NiO, FRFPMS can be readily applied to snapshots with atomic displacements generated using phonon dispersions calculated by DFT.

Below, we detail the methodologies employed, beginning with the TACAW-based magnon EELS simulations, followed by the FRFPMS phonon EELS calculations.

**Magnon EELS simulations.** The numerical simulations were performed with the TACAW method [33]. Specifically, the momentum-resolved EELS signals were computed as the time-to-energy Fourier transform of the temperature-dependent time auto-correlation of the electron beam wave-function obtained from the magnetic (Pauli) multislice method [56]. Atomistic spin dynamics (ASD) simulations in UppASD [34] were conducted on a 16 × 16 × 96 supercell (of dimensions 6.672 nm × 6.672 nm × 40.032 nm) of NiO cubic unit cells, using the experimental parameters reported in Ref. [57]. For the spin-Hamiltonian parameters, we used the values from the bottom row of Table III in Ref. [57]. Given that we utilised a cubic cell, the nearest-neighbour exchange interaction $J_1$ was set as the average of $J_1^+$ and $J_1^-$ from the same row. To account for the effect of the microscope's objective lens on the sample environment, a 1.5 T static external magnetic field oriented along the [001] direction was included. Oxygen atoms were excluded from the ASD simulations due to their negligible magnetic moment. Employing a Gilbert damping $\alpha = 5\times10^{-4}$ and a time step of 0.1 fs, a thermalisation phase of 70,000 steps was followed by a 7.813 ps trajectory at 300 K for generating snapshots every 13 fs, enabling the exploration of magnon frequencies up to 159 meV. Note that from ASD simulations we determined that the employed parameters of the spin-Hamiltonian predict a Néel temperature of 304 K, instead of the known experimental value of 523 K. Therefore, in our calculations we employed an effective temperature $T_e$ = 174.4 K, since 300 K = $T_e$ (523/304). In total, 6002 snapshots were generated and the magnon EELS signal was obtained as the average over 115 sets of 301 consecutive snapshots mutually offset by 50 snapshots. Each set spanned 3.913 ps, providing an energy resolution of 1.06 meV. The electron-beam exit wave functions were computed using the Pauli multislice method on a numerical grid of 1344 × 1344 points with 4032 slices across the NiO supercell's thickness, including the oxygen atoms (with a zero magnetic moment). A 60 kV electron probe with a 2.25 mrad convergence semi-angle, propagating along the [001] direction, was employed. We have used the parametrised magnetic fields and vector potentials developed in Ref. [58]. The Debye-Waller factor (from Table S.V in the supplemental material of Ref. [57]) and the absorptive optical potential (see appendix B of Ref. [30]) were included to simulate, in a first approximation, the effect of phonon excitations on elastic scattering.

**Phonon EELS calculations.** We have used FRFPMS [56] with 34 frequency bins spanning a range from 0 THz up to 17 THz in 0.5 THz-wide intervals. Within each of the frequency bins, we have averaged over 128 structure snapshots. The supercell size for FRFPMS simulations was 10 × 10 × 98 cubic Bravais unit cells of NiO. For phonon EELS, a conventional multislice algorithm was used, as implemented in the DrProbe code [59], using a numerical grid of 840 × 840 × 784 pixels. Convergence semi-angle, acceleration voltage, and aperture shape were all set according to the experimental geometry.

In contrast to previous applications of the FRFPMS method, we did not use molecular dynamics to generate snapshots of the vibrating structure. Instead, we first performed density functional theory (DFT) simulations of the phonon eigenmodes (see below). Using this information, we have generated structure snapshots in an approach following Refs. [60, 61] by calculating atomic displacements due to random excitation of phonon modes following their thermal population at 300 K. However, instead of summing over all the phonon modes, we have split them by their eigen-frequencies into the above-mentioned 34 frequency bins and generated sets of 128 snapshots for each frequency bin separately. Considering the small unit cell of NiO, this approach brings DFT-level precision at a lower computational cost than training a machine-learning inter-atomic potential for subsequent molecular dynamics simulations.

**Density functional theory calculations.** DFT calculations were performed using VASP [62] at the meta-GGA level using the r$^2$SCAN functional [63] with the PAW pseudo-potentials [64] containing the kinetic-energy density of core electrons. A 2 × 2 × 2 supercell of the conventional standard Fm$\overline{3}$m unit cell of NiO was geometrically optimised; the cell shape, cell volume, and atomic positions were allowed to relax to a tolerance of 1 meV/Å, to capture the distortion away from the cubic symmetry caused by antiferromagnetic ordering along the [111] direction. For all calculations, a Γ-centred $k$-point grid with spacing 2π(0.06) $Å^{-1}$ was used with a plane-wave cut-off of 750 eV. The python package phonopy [65, 66] was used to generate the displacements required to calculate force constants in a 4 × 4 × 4 supercell of the conventional standard unit cell. The dielectric constant and Born effective charges were also calculated in a 4 × 4 × 4 supercell using the finite differences approach. These are for use in the non-analytical correction [67-69], required due to the polar nature of NiO.

Phonon modes were sampled on a grid of 5 × 5 × 49 **q**-points spanning the Brillouin zone of a 2 × 2 × 2 supercell of NiO used in DFT simulations for calculation of the force matrix. The grid was chosen in a way to guarantee that atomic displacements are periodic across the boundaries of the simulation supercell used in phonon EELS simulations (see above).

### Data availability

The data that support the findings of this study are available from the corresponding authors upon request.

### Code availability

All the codes used in this work are available from the corresponding authors upon request.

### Additional References

[51] Krivanek, O. L. *et al.* Progress in ultrahigh energy resolution EELS. *Ultramicroscopy* **203**, 60–67 (2019). https://doi.org/10.1016/j.ultramic.2018.12.006.

[52] Krivanek, O. L. *et al.* Gentle STEM: ADF imaging and EELS at low primary energies. *Ultramicroscopy* **110**, 935–945 (2010). https://doi.org/10.1016/j.ultramic.2010.02.007.

[53] Hage, F.S. *et al.*, Topologically induced confinement of collective modes in multilayer graphene nanocones measured by momentum-resolved STEM-VEELS. *Physical Review B* **88**, 155408, (2013). https://doi.org/10.1103/PhysRevB.88.155408.

[54] Loane, R. F., Xu, P., and Silcox, J. Thermal vibrations in convergent-beam electron diffraction. *Acta Crystallographica Section A: Foundations of Crystallography* **47**, 267-278 (1991). https://doi.org/10.1107/S0108767391000375.

[55] Zeiger, P. M. & Rusz, J. Efficient and Versatile Model for Vibrational STEM-EELS. *Physical Review Letters* **124**, 025501 (2020). https://doi.org/10.1103/PhysRevLett.124.025501.

[56] Lyon, K. & Rusz, J. Parameterization of magnetic vector potentials and fields for efficient multislice calculations of elastic electron scattering. *Acta Crystallographica Section A: Foundations and Advances* **77**, 509–518 (2021). https://doi.org/10.1107/S2053273321008792.

[57] Hutchings, M. T. & Samuelsen, E. J. Measurement of spin-wave dispersion in NiO by inelastic neutron scattering and its relation to magnetic properties. *Physical Review B* **6**, 3447–3461 (1972). https://link.aps.org/doi/10.1103/PhysRevB.6.3447.

[58] Edström, A., Lubk, A., and Rusz, J. Elastic Scattering of Electron Vortex Beams in Magnetic Matter. *Phys. Rev. Lett.***116**, 127203 (2016). https://doi.org/10.1103/PhysRevLett.116.127203.

[59] Barthel, J. Dr. Probe: A software for high-resolution STEM image simulation. *Ultramicroscopy* **193**, 1–11 (2018). https://doi.org/10.1016/j.ultramic.2018.06.003.

[60] Muller, D. A., Edwards, B., J. Kirkland, E. & Silcox, J. Simulation of thermal diffuse scattering including a detailed phonon dispersion curve. *Ultramicroscopy* **86**, 371–380 (2001). https://doi.org/10.1016/S0304-3991(00)00128-5.

[61] Chen, X., Kim, D. S. & LeBeau, J. M. A comparison of molecular dynamics potentials used to account for thermal diffuse scattering in multislice simulations. *Ultramicroscopy* **244**, 113644 (2023). https://doi.org/10.1016/j.ultramic.2022.113644.

[62] Kresse, G. & Furthmüller, J. Efficient iterative schemes for ab initio total-energy calculations using a plane-wave basis set. *Physical Review B* **54**, 11169–11186 (1996). https://link.aps.org/doi/10.1103/PhysRevB.54.11169.

[63] Furness, J. W., Kaplan, A. D., Ning, J., Perdew, J. P. & Sun, J. Accurate and Numerically Efficient $r^2$SCAN Meta-Generalized Gradient Approximation. *The Journal of Physical Chemistry Letters* **11**, 8208–8215 (2020). https://doi.org/10.1021/acs.jpclett.0c02405.

[64] Kresse, G. & Joubert, D. From ultrasoft pseudopotentials to the projector augmented-wave method. *Physical Review B* **59**, 1758–1775 (1999). https://doi.org/10.1103/PhysRevB.59.1758.

[65] Togo, A. First-principles Phonon Calculations with Phonopy and Phono3py. *Journal of the Physical Society of Japan* **92**, 12001 (2023). https://doi.org/10.1103/PhysRevB.59.1758.

[66] Togo, A., Chaput, L., Tadano, T. & Tanaka, I. Implementation strategies in phonopy and phono3py. *Journal of Physics: Condensed Matter* **35**, 353001 (2023). https://dx.doi.org/10.1088/1361-648X/acd831.

[67] Gonze, X. & Lee, C. Dynamical matrices, Born effective charges, dielectric permittivity tensors, and interatomic force constants from density-functional perturbation theory. *Physical Review B* **55**, 10355–10368 (1997). https://doi.org/10.1103/PhysRevB.55.10355.

[68] Gonze, X., Charlier, J.-C., Allan, D. C. & Teter, M. P. Interatomic force constants from first principles: The case of α-quartz. *Physical Review B* **50**, 13035–13038 (1994). https://doi.org/10.1103/PhysRevB.50.13035.

[69] Lee, S. *et al.* Magnetoelastic coupling forbidden by time-reversal symmetry: Spin-direction-dependent magnetoelastic coupling on MnO, CoO, and NiO. *Physical Review B* **93**, 064429 (2016). https://link.aps.org/doi/10.1103/PhysRevB.93.064429.

**Acknowledgements.** SuperSTEM is the National Research Facility for Advanced Electron Microscopy supported in part by the Engineering and Physical Sciences Research Council (EPSRC) under grant number EP/W021080/1. We acknowledge further financial support from the EPSRC via grants number EP/V048767/1, EP/Z531194/1, EP/V036432/1, as well as the Royal Society via grant no. IES/R1/211016. We acknowledge the Swedish Research Council (grant no. 2021-03848), Olle Engkvist's foundation (grant no. 214-0331), STINT (grant no. CH2019-8211), Knut and Alice Wallenberg Foundation (grant no. 2022.0079), and eSSENCE for financial support. The simulations were enabled by resources provided by the National Academic Infrastructure for Supercomputing in Sweden (NAISS), partially funded by the Swedish Research Council through grant agreement no. 2022-06725.

**Author contributions.** DMK and QMR designed, performed experiments and analysed the experimental results. JCI analysed the experimental results. JACR, PZ, AB and JR designed and performed magnon calculations. AK, JdN, JR, PZ, BM and VL designed and performed phonon calculations. KEH prepared samples for analysis. All authors contributed equally to the analysis, interpretation and preparation of the manuscript.

**Competing interests.** The authors declare no competing interests.

## Extended Data Figures

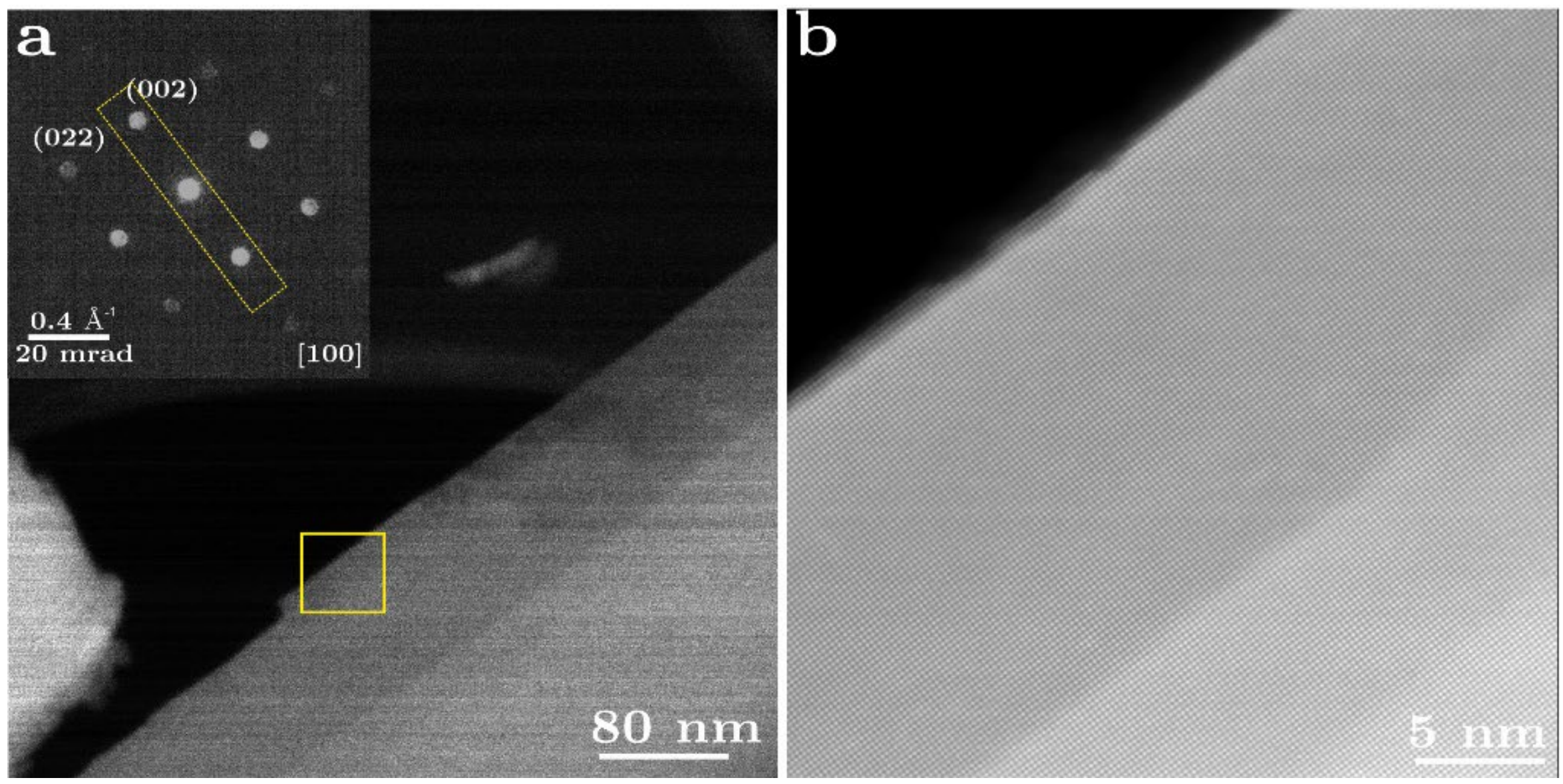

**Extended Data Figure 1 | Imaging of the NiO single-crystal sample.** ***a.*** *Annular dark-field image of the edge of a NiO single crystal, acquired with a 2.25 mrad convergence angle (~1.3 nm probe) along the* [100] *zone axis. Inset: experimental diffraction pattern along the NiO* [100] *zone axis at a 2.25 mrad convergence angle, with the monochromating slit inserted, showing the orientation of the EELS collection slot aperture along the (002) row of reflections (dashed yellow box).* ***b.*** *Atomic-resolution high-angle annular-dark-field STEM (HAADF-STEM) image acquired using a 31 mrad convergence angle from the area marked with a yellow rectangle in panel a.*

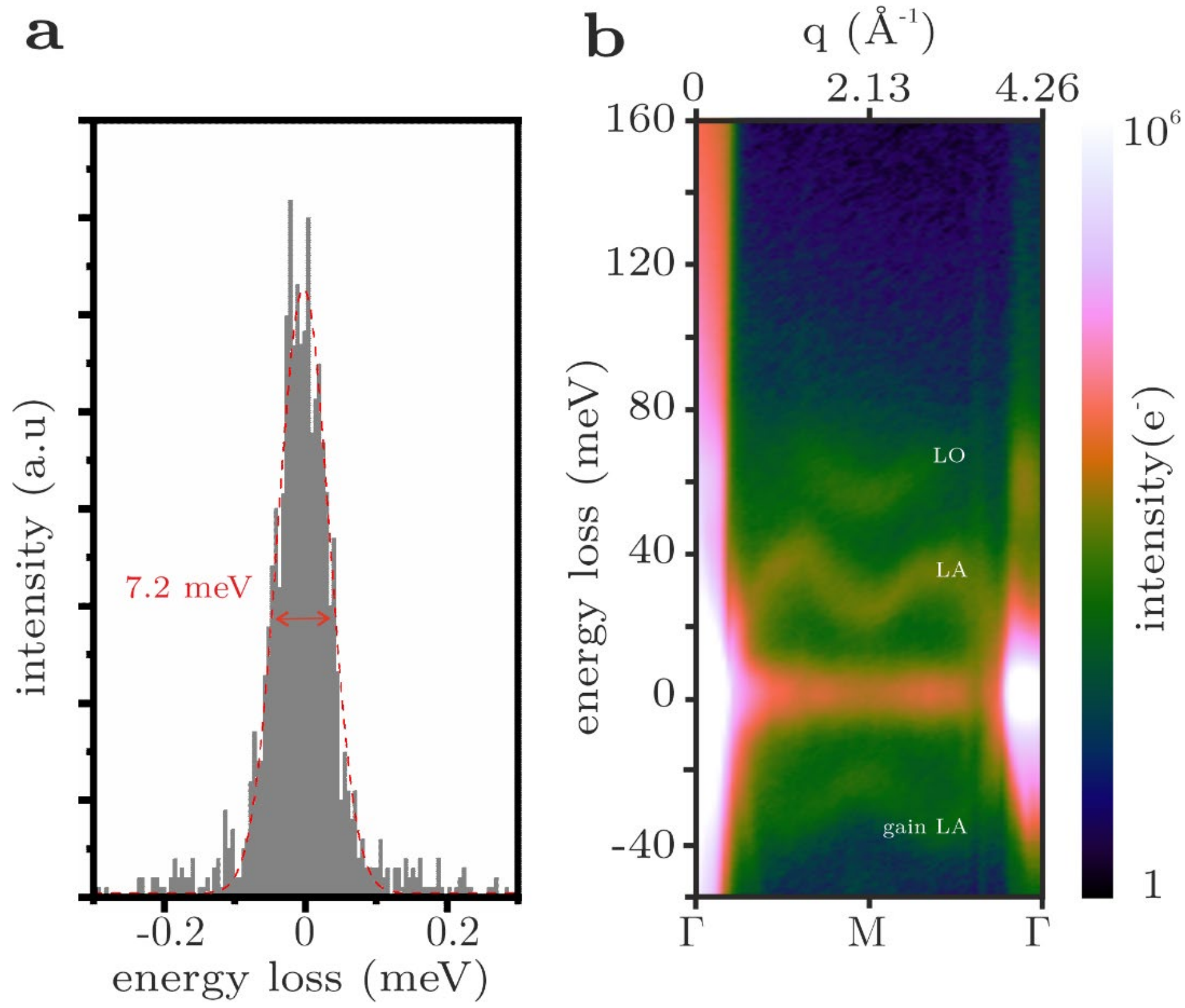

**Extended Data Figure 2 | Vibrational EELS measurements of NiO.** ***a.*** *EELS spectrum corresponding to a single acquisition frame (75 ms) in vacuum, showing a ZLP measuring 7.2 meV at the FWHM.* ***b.*** *As-acquired* $\omega$-**q** *maps along the 220 row of NiO reflections, displaying the dispersion of the NiO LA / LO phonon branches, as well as the LA gain branch, presented on a logarithmic intensity scale. The dataset corresponds to 15,000 integrated frames (75 ms each).*

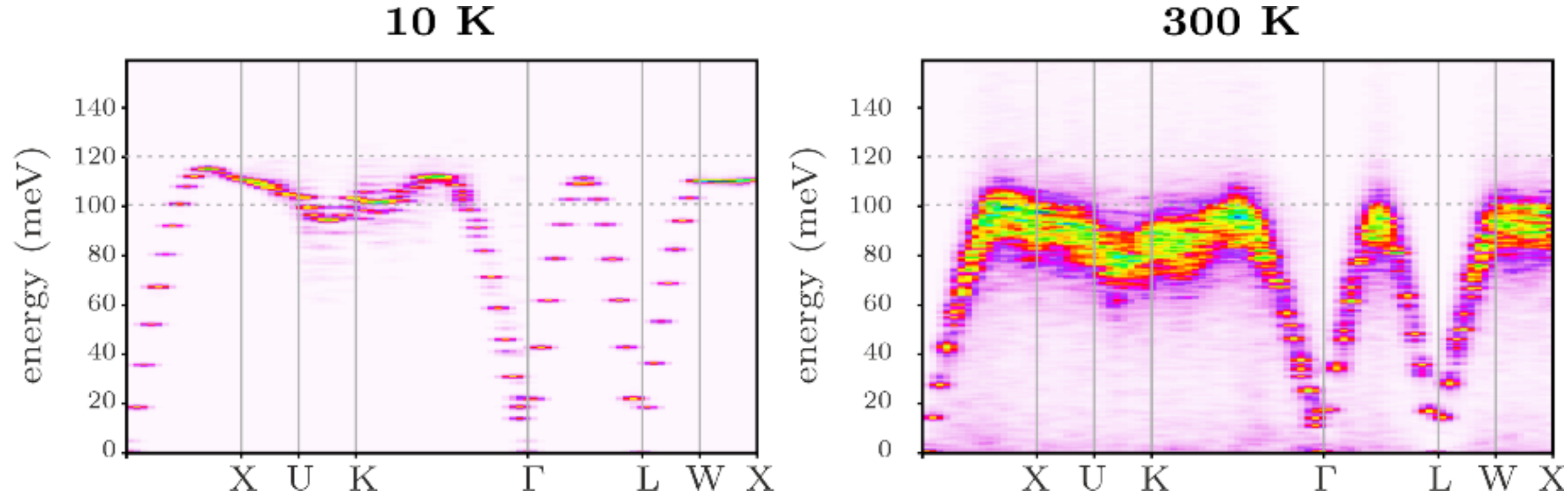

**Extended Data Figure 3 | Magnon dispersions at different temperatures.** *Calculated magnon dispersions corresponding to the dynamical structure factor computed using UppASD [34] for a NiO supercell consisting of 32 × 32 × 32 repetitions of the cubic unit cell with periodic boundary conditions applied in all directions. The atomistic spin dynamics simulations were performed employing a time step of 0.1 fs over a total simulation time of 15 ps with the remaining parameters the same as the main text.*

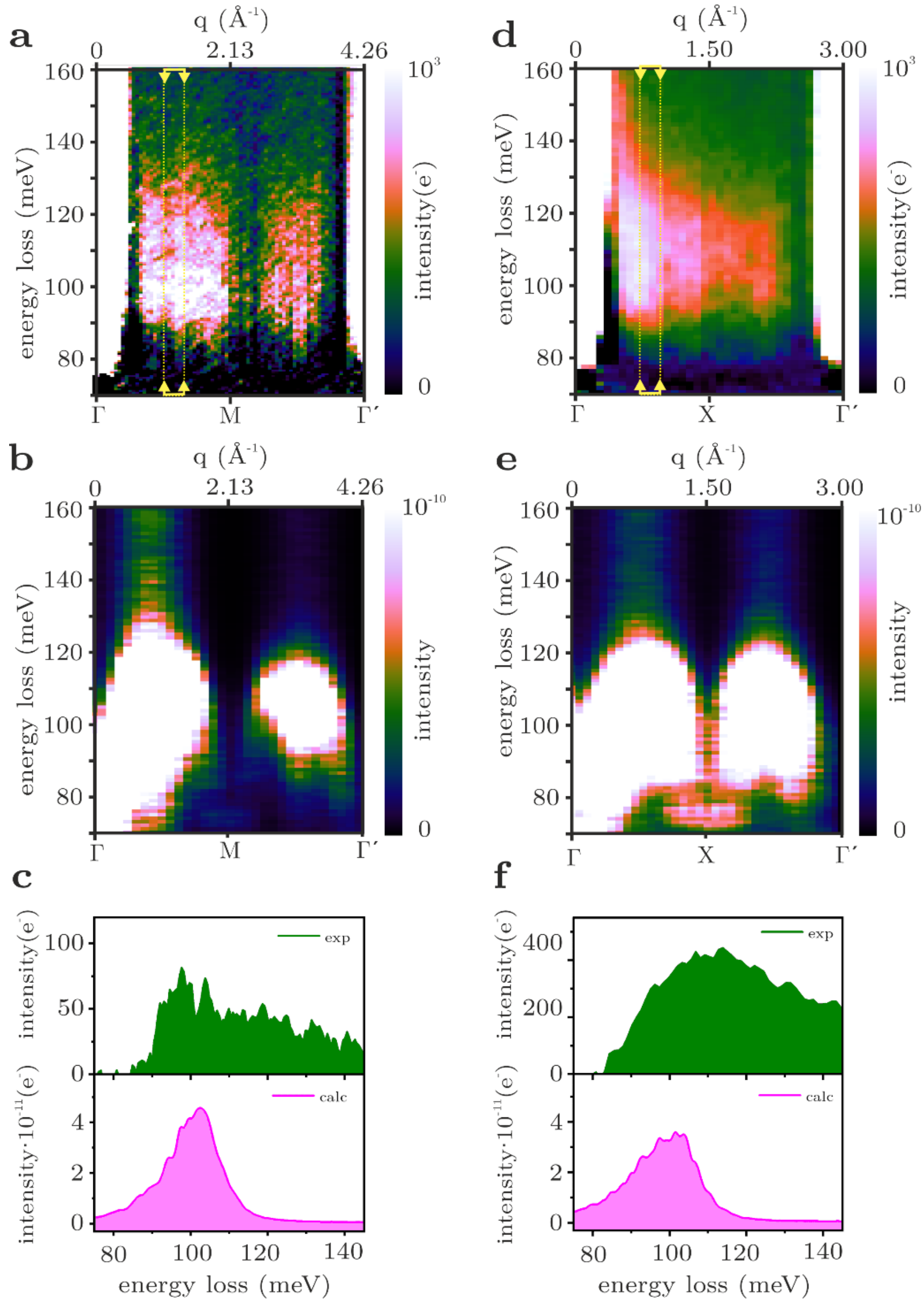

**Extended Data Figure 4 | Experiment *vs.* Theory.** *Experimental background-subtracted (***a,d***) and calculated (***b,e***) ω-***q*** EELS maps showing the dispersion of the magnon bands above 100 meV.* **c,f.** *Extracted spectra over a narrow momentum window (Δq = 0.22 Å$^{-1}$, to avoid spectral broadening through momentum averaging, at the value at which the intensity is maximum along Γ → M (q = 1.24 Å$^{-1}$) and Γ → X (q = 0.97 Å$^{-1}$), marked by yellow arrows and dashed lines in Extended Data Fig. 4a and 4d, respectively. Calculated spectra at the same wave-vector and averaged over a similar momentum window highlight the shape of the peaks, with a rising edge from 80 meV reaching a maximum at ~100 meV, before a weaker feature extending up to 120 meV, with the peak observed along Γ → X being broader and more rounded.*

# Supplementary Information

# Magnon spectroscopy in the electron microscope

Demie Kepaptsoglou[1,2,3*†], José Ángel Castellanos-Reyes[4†], Adam Kerrigan[2,3], Júlio Alves do Nascimento[2,3], Paul M. Zeiger[4], Khalil El hajraoui[1,2], Juan Carlos Idrobo[5,6], Budhika G. Mendis[7], Anders Bergman[4], Vlado K. Lazarov[2,3], Ján Rusz[4*], and Quentin M. Ramasse[1,8,9*]

*[1]SuperSTEM Laboratory, SciTech Daresbury Campus, Daresbury, WA4 4AD, UK.*

*[2]School of Physics, Engineering and Technology, University of York, Heslington, YO10 5DD, UK.*

*[3]JEOL NanoCentre, University of York, Heslington, YO10 5DD, UK.*

*[4]Department of Physics and Astronomy, Uppsala University, Box 516, Uppsala, 75120, Sweden.*

*[5]Materials Science and Engineering Department, University of Washington, Seattle, WA 98195, USA.*

*[6]Physical and Computational Sciences Directorate, Pacific Northwest National Laboratory, Richland, WA 99354, USA.*

*[7]Department of Physics, Durham University, Durham, DH1 3LE, UK.*

*[8]School of Chemical and Process Engineering, University of Leeds, Leeds, LS2 9JT, UK.*

*[9]School of Physics and Astronomy, University of Leeds, Leeds, LS2 9JT, UK.*

*Corresponding author(s). E-mail(s): dmkepap@superstem.org; jan.rusz@physics.uu.se; qmramasse@superstem.org.

†These authors contributed equally to this work.

## Supplementary Note 1: Intensity scaling

Due to the $1/E$ dependence of the classical model for the harmonic oscillator strength (where $E$ is the energy of the oscillation) [70], phonon scattering data is often displayed with the observed experimental intensity scaled by the energy loss $E$. This offers a more direct comparison to *ab initio* calculations (see, e.g., neutron inelastic scattering data in ref. [71]). Accounting for the energy-dependent occupancy of phonon excitations with a Bose-Einstein model (assuming zero chemical potential), an additional factor of $1/E$ appears in the oscillator strength model when approximating $1/[exp(E/k_BT) - 1] \sim 1/E$, if $E << k_BT$, where $k_B$ is the Boltzmann constant, and $T$ is the temperature. As a result, a total scaling of the experimental scattered intensity by $E^2$ can be justified at low energy losses.

A number of research groups have therefore proposed displaying high-energy-resolution EELS data after scaling by the square of the energy loss, that is, displaying [intensity × (energy-loss)$^2$] *vs.* (energy-loss) [72]. This scaling by $E^2$ should strictly only be valid at energies well below $k_BT$, *i.e.* a few meV at room temperature. Nevertheless, even when applied across larger energy-loss windows, this strategy provides a useful means to enhance the visibility of weaker intensity features above the decaying zero-loss-peak (ZLP) tail in the meV range, while avoiding possible errors and subjectivity in background fitting. Considered as a pure 'data scaling' strategy (as would be a logarithmic intensity display), its helps to visualise signals with vastly different intensity levels on the same panel, including in close vicinity to the ZLP. Supplementary Fig. 1 illustrates the effectiveness of this data scaling approach in revealing more clearly the presence of the magnon scattering branches above the phonon energy range in the full $\omega$-**q** maps for NiO.

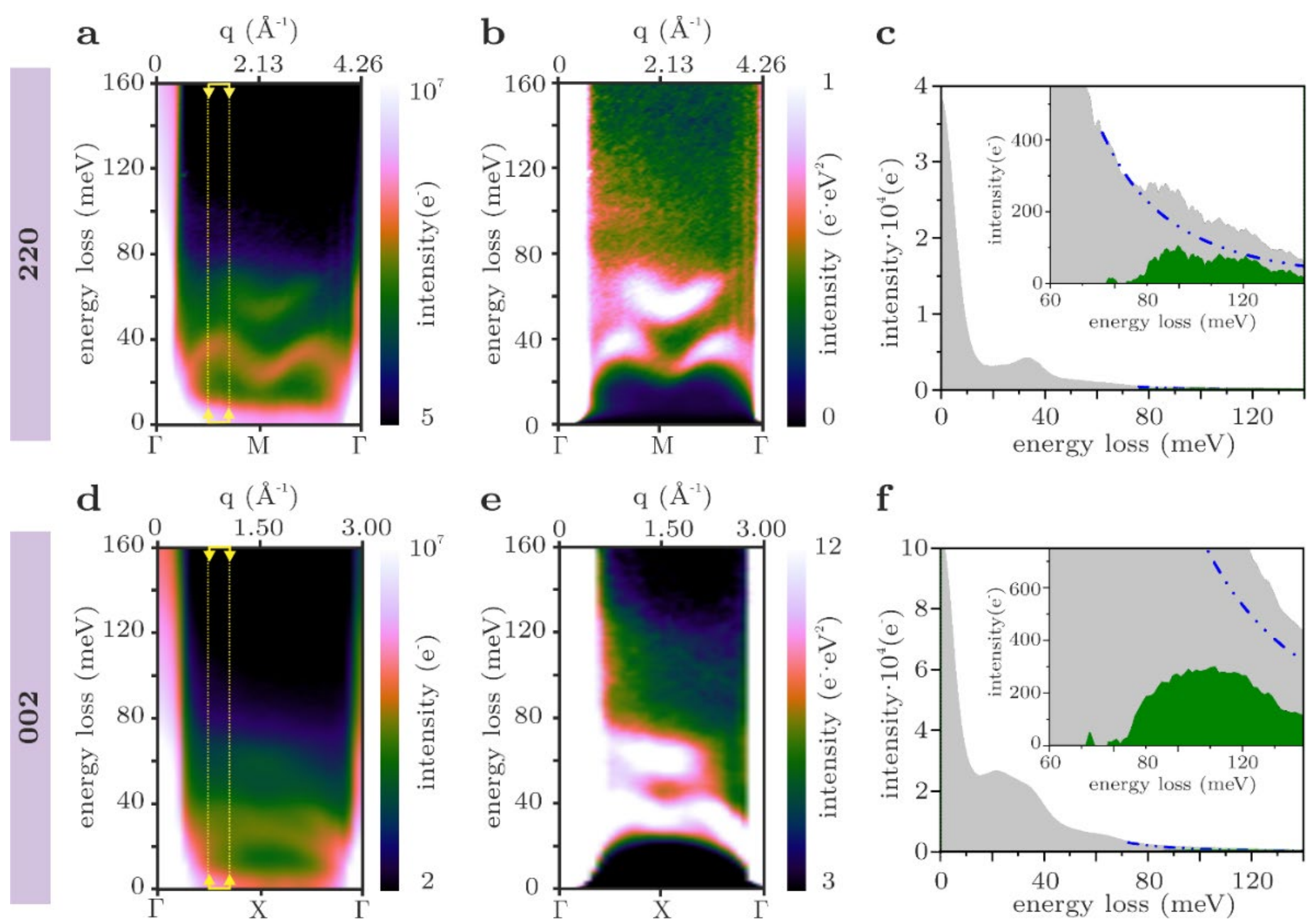

**Supplementary Fig. 1 | Momentum-resolved vibrational EELS measurements of NiO.** ***a,d.*** *As-acquired ω-***q** *maps along the 220 and 002 rows of reflections, respectively, showing the dispersion of the NiO LA and LO phonon branches, displayed on a logarithmic intensity scale.* ***b,e.*** *ω-***q** *maps along the same reciprocal space directions scaled by the square of the energy loss,* $E^2$*.* ***c,f.*** *Integrated spectra at the specific momentum positions marked by arrows in panels* ***a,d****. Inset: spectral intensity (green shaded areas) after background subtraction (the blue dotted lines illustrate the background model, while the shaded grey area is the unprocessed intensity).*

## Supplementary Note 2: Background subtraction

The use of a power-law function to remove any decaying background is one of the most common approaches in EELS, as it models population statistics quite effectively and appears to be relatively robust, even in the vibrational EELS range [73, 74]. However, the stability of the model across neighbouring pixels in 2-dimensional datasets like $\omega$-**q** maps can be poor when the energy fitting window is narrow. This is often the case for very low energy losses close to the zero-loss peak tail, or for signals superimposed on the tail of other losses close in energy. In turn, this can lead to subjective results depending on a given user's fitting window selection. As a result, background subtraction is often employed on a case-by-case basis, considering various background options, which can sacrifice physical justification to improve visibility [75].

Here, we explore different fitting options for the removal of the LO phonon decaying background immediately before the energy range where our simulations predict the presence of the magnon signal. Supplementary Figure 2 presents the $\omega$-**q** maps along the 220 systematic rows of reflections, using first-order logarithmic-polynomial and power-law models for comparable energy windows. In both cases, the resulting signal shows a remarkable qualitative resemblance to the calculated curves in Figure 3 of the main manuscript, demonstrating the robustness of the signal against the choice of background model. However, the use of the power-law background appears to yield noisier maps, as it is more susceptible to variations in phonon intensity tail across the narrow fitting energy window, resulting in some negative values, particularly on the weaker M→Γ' magnon branch.

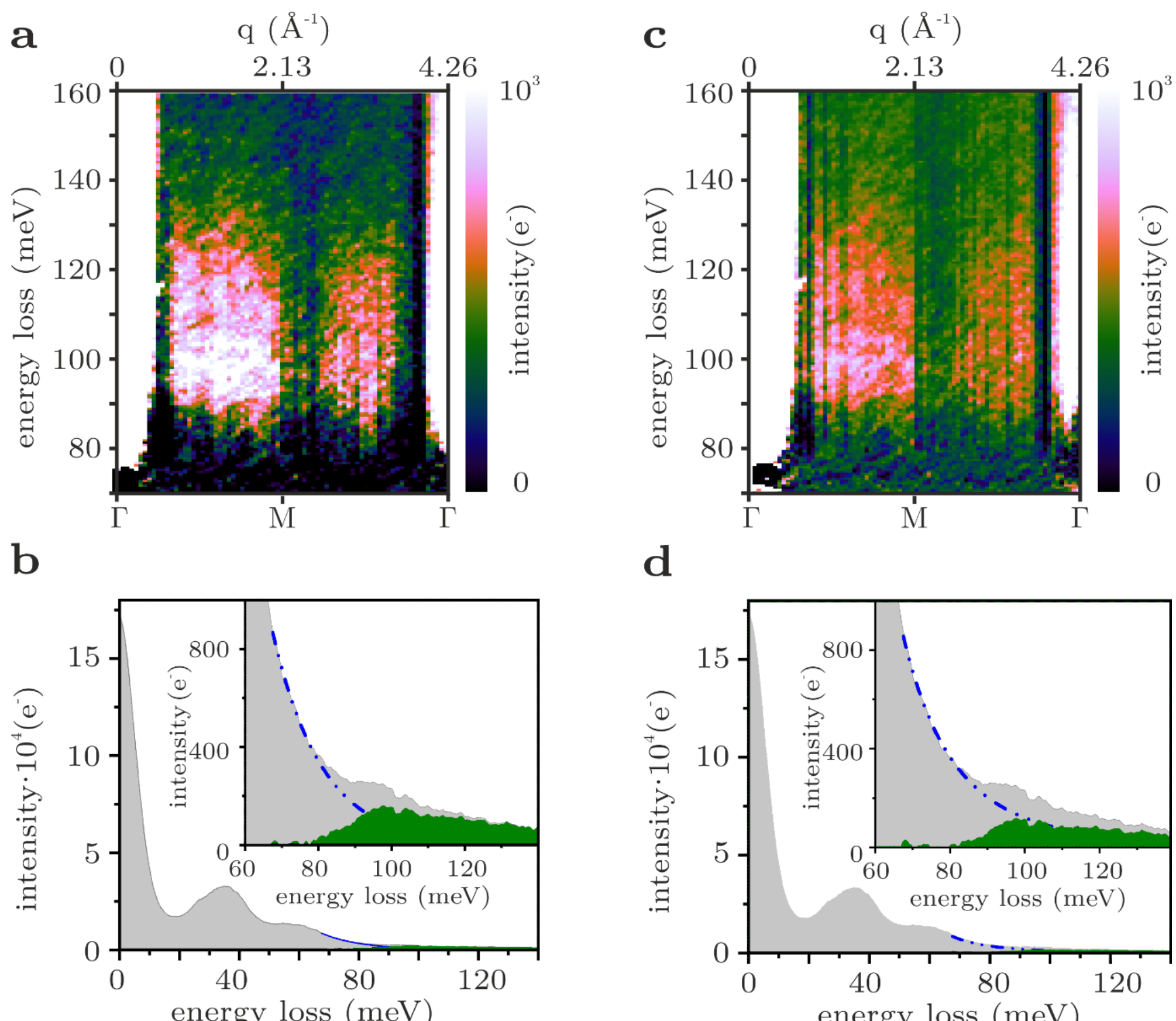

**Supplementary Fig. 2 | Robustness of magnon signal against background-subtraction models.** *Background-subtracted ω-***q** *maps along Γ→M→Γ' and corresponding extracted spectra (inset: detail after background subtraction, green shaded area; the dotted blue lines are the decaying background model, with the shaded grey area is the unprocessed spectral intensity) using:* ***(a,b)*** *a first-order log-polynomial background model and* ***(c,d)*** *a power-law model.*

**Supplementary Note 3: Data accumulation and signal-to-noise**

As discussed in the Methods section of the paper, the $\omega$-**q** EELS datasets were acquired as multiple frames covering the entire two-dimensional extent of the spectrometer's camera. A typical single dataset comprised 15,000 frames (limited by data file size), which were subsequently aligned using rigid image registration to account for possible instabilities, and integrated. The final data presented are sums of multiple such integrated datasets acquired consecutively at the exact same sample position, on the same day and under identical experimental conditions, with no other post-processing applied. The signal accumulation was used to improve the signal-to-noise ratio and the visibility of the inherently weak magnon signal in the final dataset.

Datasets were not averaged over several days' acquisition, over series acquired at slightly different locations, or across datasets involving other experimental changes, to maintain the nanometre scale resolution of the experiments on the very same sample area.

Supplementary Figure 3 shows $\omega$-**q** maps along the 220 and 002 rows of systematic reflections; panels **a** and **c** correspond to sums of 90,000 (6 x 15,000, SNR = 22) and 60,000 (4 x 15,000, SNR = 59) individual camera frames, respectively, while panels **b** and **d** consist of partial frame-integrated datasets from **a** and **c,** comprising only the first 30,000 (SNR = 10.7) and 15,000 (SNR = 16) frames from the respective series. The smaller number of frames required to obtain comparable SNR in the case of the 002 data is due to a tip change occurring between the two sets of experiments, resulting in higher probe currents available at otherwise similar electron optical settings. The magnon dispersion bands and their separation in the momentum direction are clearly discernible in both partial datasets, albeit with higher noise levels as quantified by the calculated SNRs, which remain well above 10 even with less data accumulation.

This illustrates the effect of large signal series accumulation, with a need to balance higher overall SNR with the potential for partial smearing of the final signal fine structure, due to unintended experimental instabilities through the hours-long acquisitions, that can be challenging to mitigate in post-processing. This is particularly evident along the Γ → X **q**-path (200), where the magnon intensity lobes in the dispersion band are clearly separated in panel **d**, with the intensity dropping to zero near X (as predicted by simulations). In contrast, the lobe separation and momentum variation are less evident in the larger dataset averaged over more frames, shown in panel **c**; however, this dataset has higher signal-to-noise and signal-to-background, providing an unambiguous confirmation of the presence of spectral intensity in the relevant energy window.

Here, the signal-to-noise ratio (SNR) of a given magnon dispersion diagram is defined as $SNR = \frac{\mu(d\omega,dE)}{\sigma(d\omega,dE)}$ where $\mu(d\omega, dE)$ is the mean spectral intensity in the magnon energy window after background subtraction, in otherwise unprocessed spectral data. The noise $\sigma(d\omega, dE\ )$ is in turn is estimated as the standard deviation of the recorded signal in an energy window of the same size, beyond any predicted magnon or phonon contribution (above 200 meV).

For signals whose distribution follows Poisson statistics, a more exact mathematical definition of the SNR from counting statistics can be used. While the EELS detector used here is not strictly a pure 'electron counter', its noise characteristics are known to be nearly Poisson-limited [31] and thus the following calculations were also taken into consideration for completeness (and the pixel-wise definition of the standard deviation below as the main source of counting error was used to derive the confidence bands displayed in Figure 2 or the main manuscript).

- $\mathrm{SNR} = \frac{\mathrm{M(d\omega,dE)}}{\sqrt{\mathrm{N(d\omega,dE)}}}$, where $\mathrm{N(d\omega, dE)}$ and $\mathrm{M\ (d\omega, dE)}$ are the mean spectral intensities in the magnon energy window before and after background subtraction, respectively, in otherwise unprocessed spectral data.
- $\mathrm{SNR} = \frac{\mathrm{IM(d\omega,dE)}}{\sqrt{\mathrm{IN(d\omega,dE)}}}$, where $\mathrm{IN(d\omega, dE)}$ and $\mathrm{IM\ (d\omega, dE)}$ are the integrals of the spectral intensities in the magnon energy window before and after background subtraction, respectively, in otherwise unprocessed spectral data.

All these are summarised in Supplementary Table 1 and clearly demonstrate how frame accumulation results in higher SNR irrespective of the metric used, while all values suggest that the signal is clearly above noise level.

**Supplementary Table 1 | Calculated SNR for partial and complete datasets.** *The calculated values correspond to panels presented in Fig. 2 and Supplementary Figure 3.*

| | **220** | | **002** | |
|---|---|---|---|---|
| **SRN** | 30k frames | 90k frames | 15k frames | 60k frames |
| $\boldsymbol{\frac{\mu(d\omega, dE)}{\sigma(d\omega, dE)}}$ | 10.7 | 22 | 16 | 59 |
| $\boldsymbol{\frac{M(d\omega, dE)}{\sqrt{N(d\omega, dE)}}}$ | 2 | 3 | 2 | 10 |
| $\boldsymbol{\frac{IM(d\omega, dE)}{\sqrt{IN(d\omega, dE)}}}$ | 81 | 134 | 44 | 222 |

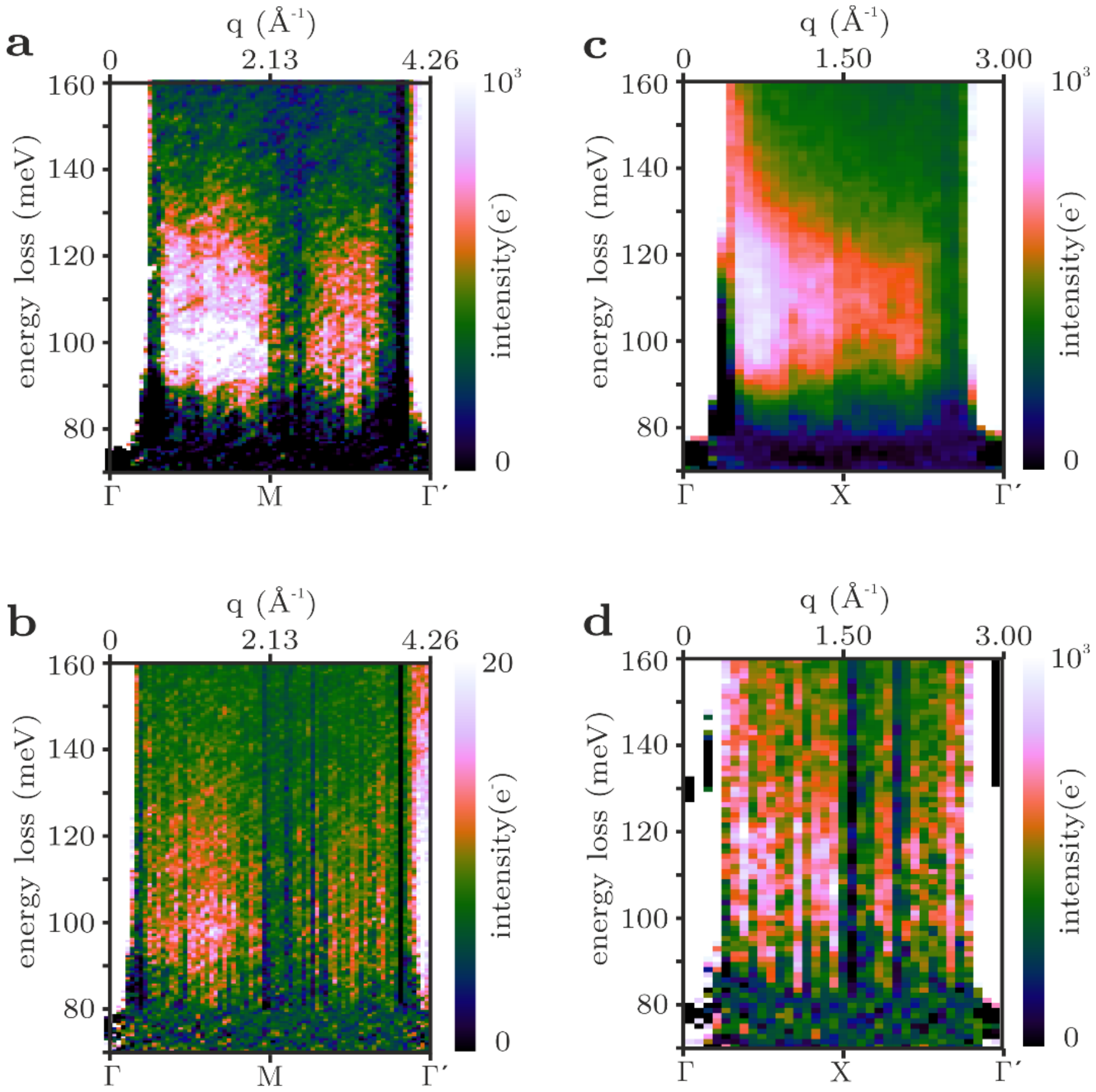

**Supplementary Fig. 3 | Frame accumulation in $\omega$-q maps along the 220 and 002 rows of reflections.** ***a**,**c.*** *$\omega$-**q** magnon maps corresponding to sums of 90,000 (SNR = 22) and 60,000 (SNR = 59) individual camera frames, respectively, corresponding to 112.5 minutes and 75 minutes of live-beam acquisition respectively (the effective experimental time is longer to account for data transfer of partial datasets between bursts of 15,000 fames).* ***b,d.*** *Partial datasets from **a** and **c** comprising 30,000 (SNR = 10.7) and 15,000 (SNR = 16) frames only (corresponding to live-beam times of 37.5 minutes and 18.75 minutes), respectively.*

**Supplementary Note 4: Experimental broadening**

The impact of finite experimental resolution in momentum and energy on the separation of the magnon dispersion bands was illustrated by numerically broadening the simulated magnon EELS dispersion diagrams along the $\Gamma \rightarrow M$ and $\Gamma \rightarrow X$ **q**-paths to match experimental limitations. This was achieved by convolution with a top-hat function of 0.7 $Å^{-1}$ width along the momentum axis (mimicking the momentum-selecting slit), and with a Gaussian of 11 meV full-width at half-maximum along the energy-dispersive axis (reflecting the effective energy resolution of the experiment in the 002 case, with the probe going through the sample after integration of 90,000 frames).

As expected, due to the shorter reciprocal space distance, the broadening has more apparent impact on the data acquired along the $\Gamma \rightarrow X$ **q**-path; the magnon peaks appear to merge into a more continuous lobe of intensity, reflecting the experimental data where, as discussed in the main manuscript, the two peaks are hard to separate on either side of the X-point.

The match in the $\Gamma \rightarrow M$ direction, where the Brillouin zone vertices are spaced further apart so the magnon lobes are still resolved, is also more 'pleasing' visually (albeit less defined) after this forward convolution.

We note that the theoretical framework for magnon-EELS simulations is still being actively developed. At present, these approaches do not yet incorporate magnon-phonon coupling or fully capture the complex influence of phononic backgrounds. A fully coupled magnon-phonon framework, which is not yet available, would be required to reproduce the complete spectral landscape, including the intricate interplay between signal and background, as well as to address effects such as background subtraction, noise, and broadening in a more systematic and quantitative way. Nonetheless, the uncoupled simulations presented here, especially after forward convolution to mimic experimental broadening, enable a clear identification of magnonic features based on their dispersion behaviour and alignment with Brillouin zone boundaries and high-symmetry points—providing compelling evidence for the interpretation of the experimentally observed excitations. These theoretical developments are ongoing and will be the subject of future communications and publications.

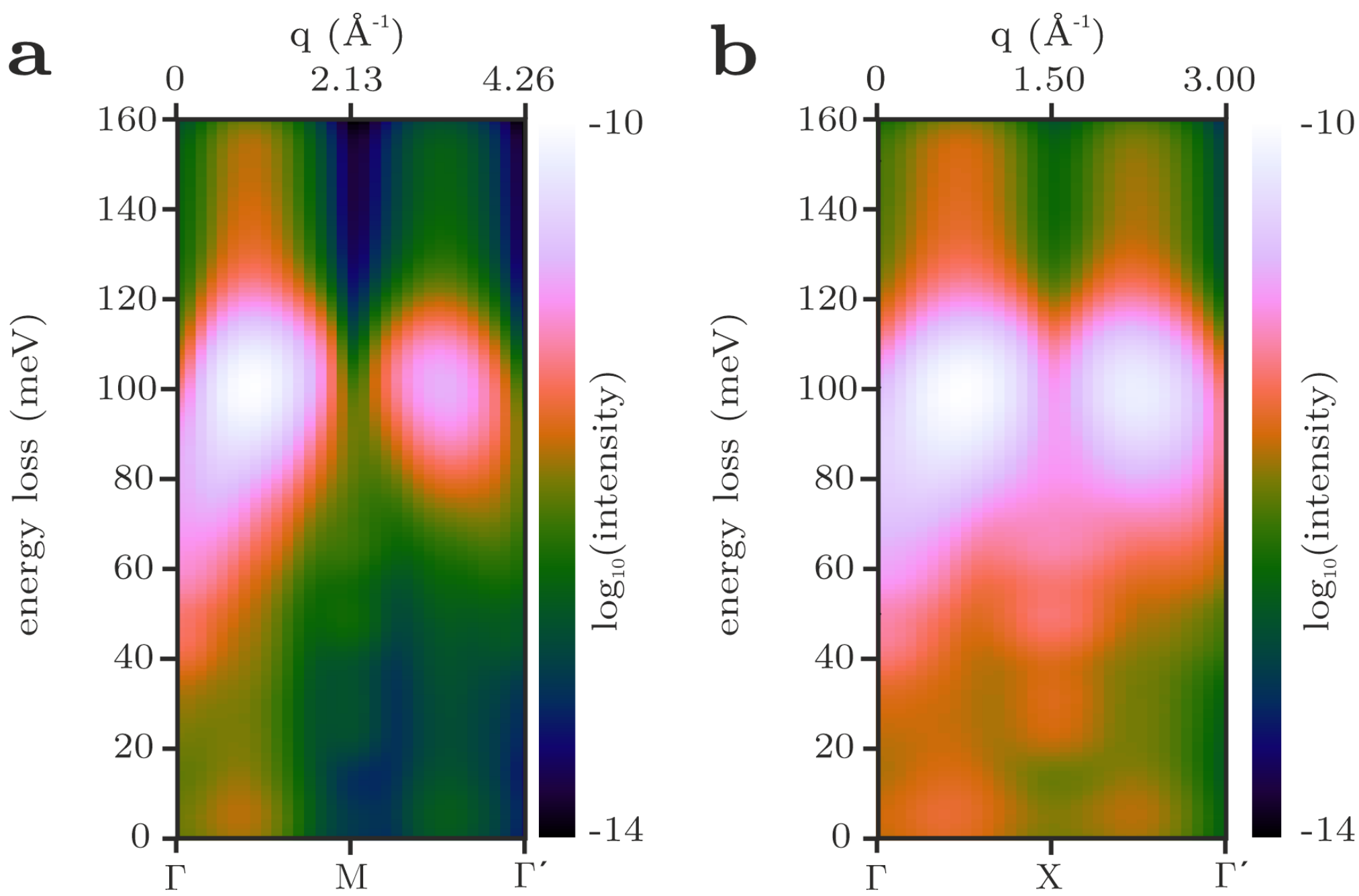

**Supplementary Fig. 4 | Broadened simulated magnon EELS dispersions**. *Simulated* $\omega$-$\boldsymbol{q}$ *magnon EELS dispersions (intensity displayed on a logarithmic colour scale) for the Γ → M* ***(a)*** *and Γ → X* ***(b)*** *directions, respectively, after convolution with a top-hat function of 0.7 Å*$^{-1}$ *width along the momentum axis and a Gaussian of 11 meV-full-width at half-maximum along the energy-dispersive axis.*

**Supplementary Note 5: Additional datasets**

Supplementary Figure 5 shows an additional, independent dataset acquired along the Γ → M **q**-path *c*orresponding to 75,000 individual (5 x 15,000) camera frames, where both the (LA and TA) phonon (Supplementary Fig. 5a) and magnon (Supplementary Fig. 5b) bands are symmetrically resolved on either side of the Γ-point.

This demonstrates the reproducibility of the results, as the dataset was acquired on a different day, and in a different area of the sample from that shown in the main manuscript, Fig. 2. The wider momentum widow presented in Supplementary Fig. 5 highlights symmetric data on either side of Γ, but due to the wider angular (momentum) range, some asymmetric energy resolution loss due to spectrometer aberrations along the length of the momentum-selecting slit results in minor feature blurring in the Γ-M-Γ' direction, compared to Γ-M-Γ''. On the right side of the Γ-point, the phonon bands are not as well-resolved, and the magnon lobes not as clearly separated. However, the phonon and magnon bands are clearly recognisable on both sides.

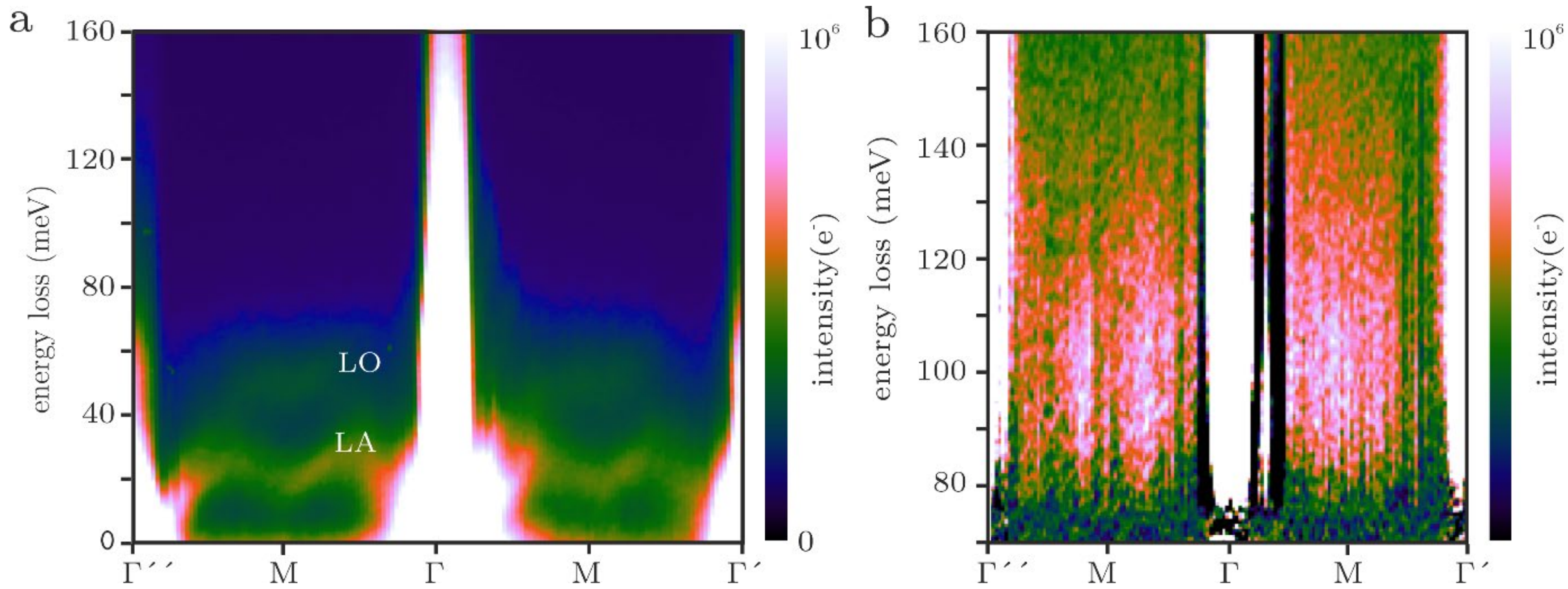

**Supplementary Fig. 5 | *ω*-q map along Γ → M q-path**. ***a.*** *ω-**q** map corresponding to 75,000 (5 x 15,000) individual camera frames.* ***b.*** *Background-subtracted magnon dispersion diagram showing a symmetric dispersion of the magnon signal across the Γ-point.*

**Supplementary Note 6: Spatially resolved spectral variations**

Spectral analysis of the spatially resolved dark-field EELS experiments shows a well-defined magnon peak at 100 meV (Supplementary Fig. 6c) throughout the NiO thin film. The intensity of this peak is significantly suppressed when the probe is positioned inside the NiO film but within a sub-nm distance of the interface with the (non-magnetic) YSZ substrate on one side, and close to the hole created immediately above the NiO thin film during the FIB sample preparation procedure on the other. The peak shape is also altered, with a slight, but noticeable broadening of the main peak near the thin film edge and the interface. Probe propagation could play a role in these observed signal changes, as mentioned in the main text in the discussion of Fig. 4. However, magnon EELS simulations for a thin slab of NiO also predict a similar effect in the vicinity of the slab's surfaces (Supplementary Fig. 7 and Supplementary Fig. 8). Although at 20 nm wide the NiO thin film is still "thick", this broadening may be linked to the appearance of confinement-related additional softer magnon modes in the ultra-thin film limit [76], a phenomenon we plan to study in future work.

Supplementary Figure 6d shows the background-subtracted NiO magnon spectrum plotted alongside spectra from the YSZ substrate and a reference spectrum acquired in identical conditions from the carbon protective layer deposited prior to FIB sample preparation. The preparation procedure had resulted in the formation of a hole (vacuum) immediately above the NiO film in the spectrum image region-of-interest, but some remaining carbon from the protective cap remained present far above the surface in other, extended-range data. This allows for a direct and self-consistent comparison.

The 'carbon' spectrum displays a characteristic, strong vibrational signature above 150 meV (attributed to C-C vibrational and C-O modes). This can also be seen in the NiO 'surface' spectrum (Supplementary Fig. 6c), as some small amount of carbon may have remained at the very surface of the film. Similarly, some faint intensity at >150 meV, *i.e.* in the range corresponding to carbon, and which could have arisen from minor contamination build-up during the hours-long acquisition, can be seen in some of the spectra in Supplementary Fig. 6c ('bulk') and 6d ('NiO' and 'YSZ'). These are distinct in energy range and spectral shape, and of such low intensity compared to any other contribution, that carbon can be excluded with confidence as a spurious contribution to the magnon signal.

YSZ does not sustain magnons, but exhibits a main optical phonon band around 76 meV corresponding to O vibrational modes [77], which we observe in our spectra and is distinct in shape and energy from the NiO magnon peak. In future work, spatially resolved dispersion diagrams across interfaces (using a small convergence angle and slit EELS aperture) [78] will provide a way to study interface-induced modifications to magnon dispersions. Here, given the non-magnetic nature of YSZ, such an experiment would simply show the individual contributions of YSZ (phonons only) or NiO (phonons and magnon).

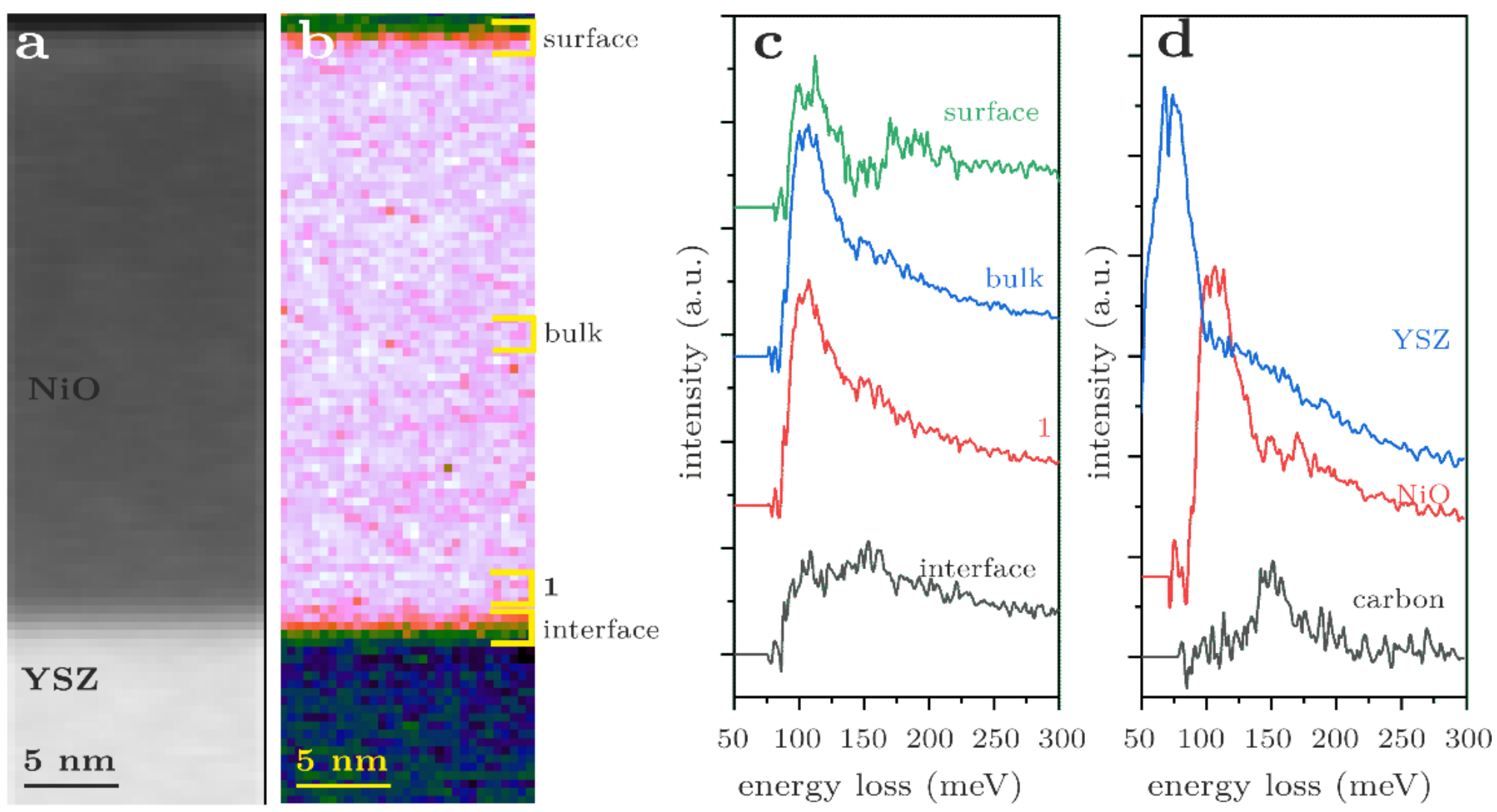

**Supplementary Fig. 6 | Spatially resolved magnon EELS measurements across a NiO thin film.** ***a.*** *Asymmetric (displaced) annular-dark-field Image (a-ADF) acquired during EELS measurements.* ***b.*** *Integrated intensity map of the magnon peak.* ***c.*** *Background subtracted magnon spectra from areas of the NiO film marked in* ***(b)****.* ***d.*** *Background-subtracted NiO magnon spectrum plotted alongside spectra from the YSZ substrate and a reference spectrum from the C protective layer of the FIB-prepared sample, respectively, as indicated.*

**Supplementary Note 7: Inelastic magnon scattering calculations for a NiO slab model**

To investigate the effects of sample geometry and electron beam positioning on magnon spectra in support of the spatially resolved data presented in Fig. 4 of the main manuscript, we performed additional magnon EELS simulations utilising the TACAW method.

Our simulations were conducted on a slab model using a supercell with 11 × 16 × 96 repetitions (of dimensions 4.587 nm × 6.672 nm × 40.032 nm) of the NiO cubic unit cell. We employed the same experimental parameters outlined in the main text for the bulk dispersion curve calculations, including a probe convergence semi-angle of 2.25 mrad. Periodic boundary conditions were applied along the [010] and [001] directions, while a vacuum boundary was introduced along the [100] direction to emulate slab geometry. Given the range of considered exchange interactions (up to 5.9 Å), a width of the slab of 4.576 nm is sufficient to prevent magnetic exchange interactions of the two surfaces as well as to reach a bulk-like behaviour in the central region. A fixed sample thickness of 40 nm was maintained for all configurations to ensure consistency.

Two distinct electron beam positions were analysed: the slab centre (wherein the beam was centred within the slab's xy-plane), and the slab's edge (wherein the beam was positioned inside the slab, 2.1 Å away from the boundary of the slab's xy-plane). The resulting spectra, depicted in Supplementary Figs. 7 and 8 are compared to the bulk NiO spectral features from the main text. Key observations include:

- **a reduced intensity at the slab's edge**. The magnon signal intensity at the edge of the slab was significantly reduced, by a factor of approximately 3-4 compared to the centre of the slab. This reduction can be attributed to the limited spatial overlap of the beam with the slab's magnonic modes as well as to local modifications of the magnon response in this region.
- **an energy redshift at the slab's edge**. Magnons near the slab's edge exhibited a redshift, consistent with localised variations in magnonic interactions influenced by boundary conditions.

These findings underscore the sensitivity of magnon spectra to both sample geometry and beam positioning. We note that due to computational limitations these EELS simulations do not fully reproduce the geometry of the spatially resolved experiments (which would have required averaging over a significantly large number of beam positions, requiring excessive computing times). Nevertheless, the simulations highlight the predicted capability of EELS to probe localised magnonic variations, emphasising its potential for studying nanoscale magnetic heterogeneities in slab systems.

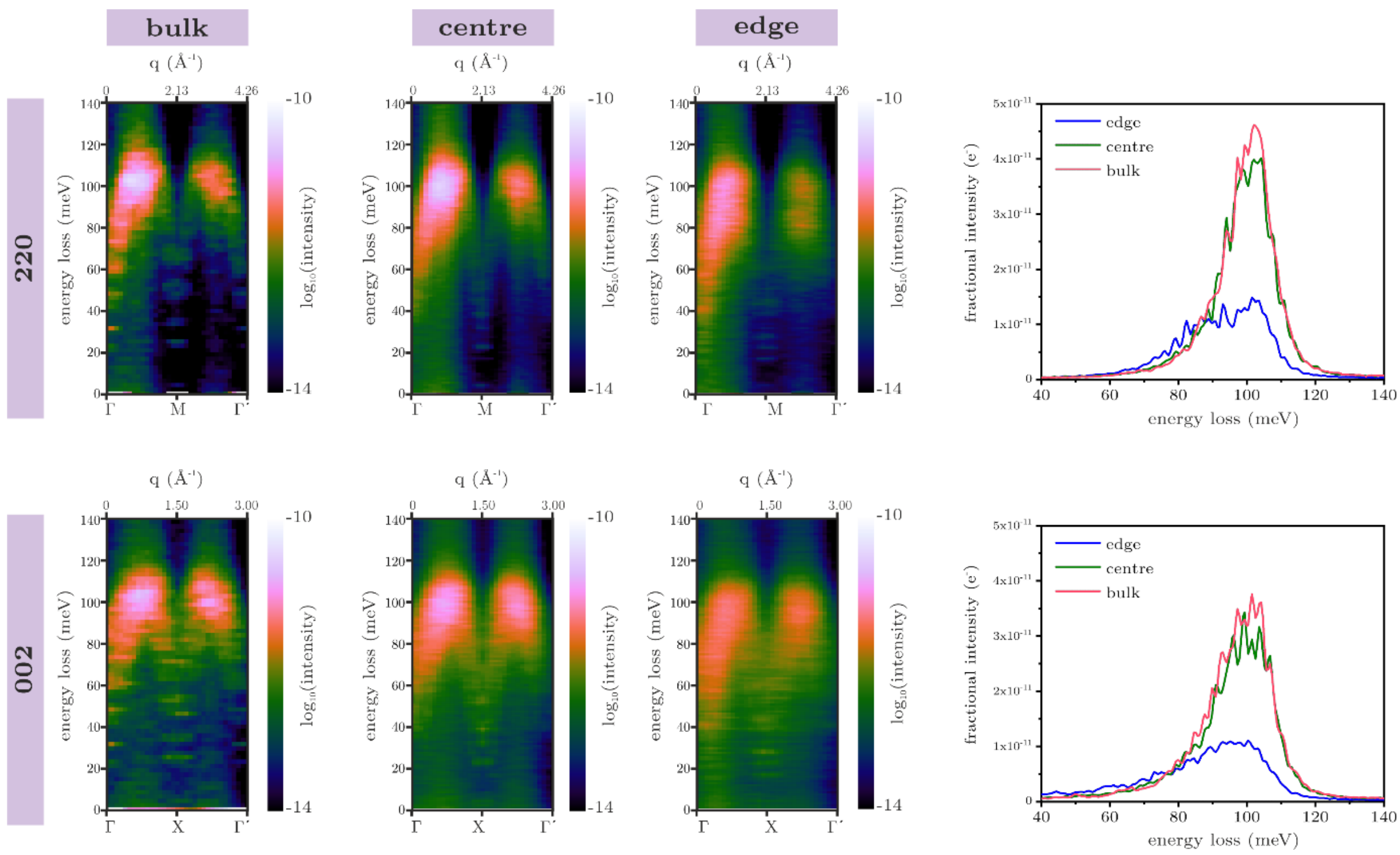

**Supplementary Fig. 7 | Spatial dependence of magnon scattering across a NiO slab.** *Simulated magnon EELS dispersions along the $\Gamma \rightarrow M$ and $\Gamma \rightarrow X$ **q**-paths of the Brillouin zone for NiO, at different positions across a 4.58 nm-wide NiO slab surrounded by vacuum. The bulk calculations, as reported in the main manuscript, Fig. 3, are reprised here and labelled 'bulk'. Integrated magnon spectra at a wave-vector of $q = 1.25$ Å$^{-1}$ show the drop of the magnon signal at the slab's edge compared to its centre, consistent with experimental observations.*

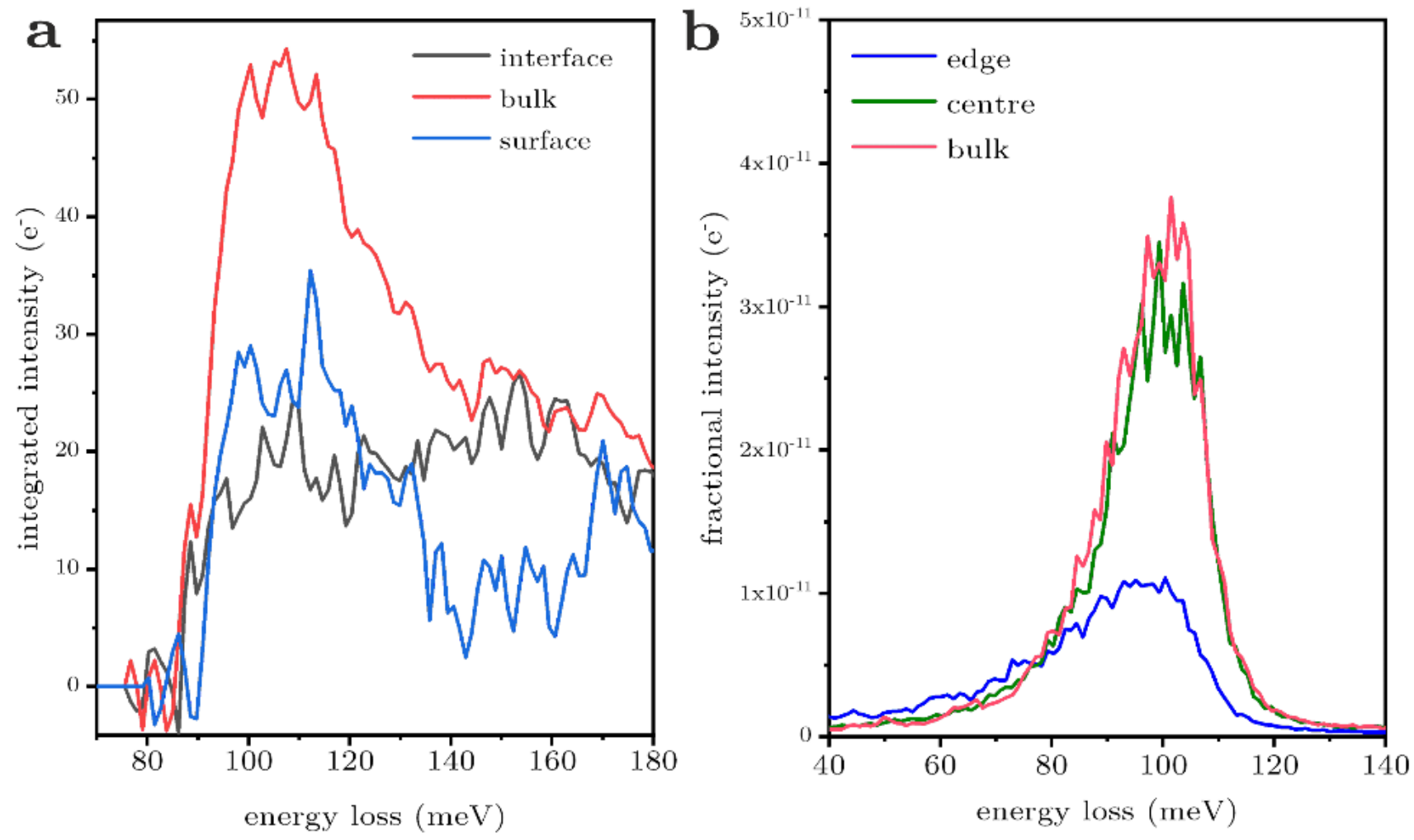

**Supplementary Fig. 8 | Comparison between experiment and theory across a NiO thin film.** ***a.*** *Background-subtracted magnon spectra from areas of the NiO film marked in Supplementary Fig. 6b.* ***b.*** *Calculated EELS magnon spectra for different positions in the NiO slab (bulk, centre, edge, as indicated).*

**Supplementary references**

[70] Sun, Q. *et al.* Mutual spin-phonon driving effects and phonon eigenvector renormalization in nickel (II) oxide. *Proceedings of the National Academy of Sciences* **119**, e2120553119 (2022). https://pnas.org/doi/full/10.1073/pnas.2120553119.

[71] Kittel, C. Introduction to Solid State Physics, 8th Edition, Wiley, New York (2005).

[72] Dellby, N., and Batson, P. Private communication.

[73] Fung, K., L., Y. *et al.* Accurate EELS background subtraction – an adaptable method in MATLAB. *Ultramicroscopy* **217**, 113052 (2020). https://doi.org/10.1016/j.ultramic.2020.113052.

[74] Haas, B. *et al.*, Atomic-Resolution Mapping of Localized Phonon Modes at Grain Boundaries. *Nano Lett.* **23**, 5975–5980 (2023). https://doi.org/10.1021/acs.nanolett.3c01089.

[75] Hachtel, J. A., Lupini, A., R., and Idrobo, J. C. Exploring the capabilities of monochromated electron energy loss spectroscopy in the infrared regime. *Scientific Reports* **8**, 5637 (2018). https://doi.org/10.1038/s41598-018-23805-5.

[76] do Nascimento, J., A. *et al.*, Confined magnon dispersions in ferromagnetic and antiferromagnetic thin films in a second quantization approach: the case of Fe and NiO. *Phys. Rev. B.* **110**, 024410 (2024). https://doi.org/10.1103/PhysRevB.110.024410.

[77] Cousland, G. P., Cui, X., Y., Ringer, S., Smith, A., E., Stampfl, A., P., J., and Stampfl, C., M. Electronic and vibrational properties of yttria-stabilised zirconia from first-principles for 10–40 mol% $Y_2O_3$. *Journal of Physics and Chemistry of Solids* **75**, 1252-1264 (2014). https://doi.org/10.1016/j.jpcs.2014.05.015.

[78] Qi, R. *et al.* Measuring phonon dispersion at an interface. *Nature* **599**, 399-403 (2021). https://doi.org/10.1038/s41586-021-03971-9.